\documentclass[aps,pre,showpacs]{revtex4}
\usepackage{bm}
\usepackage{graphicx}
\newcommand{\sump}[2]{{\sum_{#1}}\raise4pt\hbox{\hskip#2$'$}}

\begin{document}
\title{Long time limit of equilibrium glassy dynamics and replica calculation}

\author{A. Crisanti}
\email{andrea.crisanti@roma1.infn.it}

\affiliation{Dipartimento di Fisica, Universit\`a di Roma ``La Sapienza''
             and SMC, \\
             P.le Aldo Moro 2, I-00185 Roma, Italy}

\begin{abstract}
It is shown that the limit $t-t'\to\infty$ of the 
equilibrium dynamic self-energy can be computed from the $n\to 1$ limit
of the static self-energy of a $n$-times replicated system with one step
replica symmetry breaking structure. It is also shown
that the Dyson equation of the replicated system
leads in the $n\to 1$ limit to the bifurcation equation for the
glass ergodicity breaking parameter computed from dynamics.
The equivalence of the replica formalism to the long time
limit of the equilibrium relaxation dynamics is proved to all orders in 
perturbation for a scalar theory. 
\end{abstract} 


\date{2007/11/27}
\maketitle

\section{Introduction}
Spin glass and structural glasses are characterized by the presence of a 
complex structure of stable and metastable states \cite{metastable}.
The basic idea
\cite{KirThi87,CriSom92,CriHorSom93,BouMez94,MarParRit94,CriSom95,BouCugKurMez96}, 
supported  by the behavior of some mean-field spin glass models,
is that the glassy 
behavior arises because of the emergence of an exponentially large number
of metastable states that breaks the ergodicity and 
prevents the system from reaching the true thermodynamic equilibrium state. 
The logarithm of the number of 
metastable state is commonly called ``complexity'' or ``configurational'' 
entropy. The glass transition is then associated with the emergence of a 
non-zero configurational entropy. 
By extending the replica method, originally developed  for disordered systems, 
the occurrence of a non-zero configurational entropy can be conveniently
studied within the replica formalism \cite{Monasson95}.

The basic idea of the replica approach
\cite{Monasson95,MezPar99,ColMezParVer99,MezPar00,ParZam05,ParZam06} 
is to consider the equilibrium a
thermodynamics of $m$ copies of the original system interacting among them via 
an infinitesimal attractive coupling. If the free-energy landscape breaks down 
into (exponentially) many metastable states, the $m$ copies will then 
condensate into the same metastable state since their relative distances in 
such states are smaller than in the liquid (paramagnetic) phase. 
The replica method allow for an equilibrium analysis of the properties 
of the metastable states, in spite of the fact that these are originally
defined in a dynamic framework. In particular the properties of the original
system is recovered continuing the replica number $m$ to $m\leq 1$ and 
taking in the limit $m\to 1$.

If one applies the present method to mean-field $p$-spin spin glass models, 
whose low temperature phase is described by a one replica symmetry breaking 
step (1RSB), the result agrees with that from a dynamical study.
We observe that in this case the number $m$ of replicas corresponds to
the break point of the 1RSB solution of the conventional replica approach, and
hence the $m\to 1$ limit clearly corresponds to the onset of the glass phase.

All checks of
the equivalence between the replica approach and the dynamical approach
we are aware of \cite{GroKraTarVio02,WesSchWol03,MiyRei05,AndBirLef06}, 
use some approximations to compute the relevant correlation 
functions of the many-body problem. These leave open the possibility
that the equivalence is just a consequence of the limited number of 
dynamical diagrams considered in the various approximations.

To our knowledge the equivalence of two approaches for a general system is not 
yet proved. The proof of this statement is the main result reported in this 
paper.
To be more specific,
in this paper we show under rather general assumptions, that the long time
limit $t-t'\to\infty$ of the equilibrium two-times correlation function
of a glassy system can be computed from the $n\to 1$ limit of the static 
theory of an $n$-times replicated system with a 1RSB structure.
The proof is based,
following similar works \cite{Naketal83,Gozzi83} on the connection between 
static and equilibrium dynamics,  on a suitable summation 
of classes of dynamical  diagrams
of the dynamic perturbation theory generated by the Langevin dynamics, 
and showing that in the limit $t-t'\to\infty$ these approaches those of the 
$n\to 1$ limit of the equilibrium theory of an $n$-times replicated system
with a 1RSB structure.

The reader will not find in this paper the physical consequences that can be 
obtained for specific systems or models by using the results reported here. 
For these the interested reader is referred to, e.g., 
Ref. [\onlinecite{GroKraTarVio02,WesSchWol03,MiyRei05,AndBirLef06}] or to 
the forthcoming paper \cite{BirCriprep} where these will be applied 
to a simple toy system.

To keep the notation as simple as possible we shall consider only scalar fields.
The generalization to more complex fields, e.g., vector field, is 
straightforward.
Our starting point is hence the Langevin equation of the form
\begin{equation}
\label{eq:lang}
\frac{\partial\varphi(t)}{\partial t} = 
                  -\frac{\delta H[\varphi]}{\delta \varphi(t)}
                            + \eta(t)
\end{equation}
which describes the purely relaxation dynamics towards equilibrium of the 
scalar stochastic field $\varphi$ in presence of the stochastic force $\eta$.
In general both $\varphi$ and $\eta$ can be functions of both time and
space coordinates however, since we are interested into the time behavior
of correlations, space coordinates can be safely neglected. As a consequence
we shall drop any explicit space dependence of fields. The Hamiltonian 
$H[\varphi]$ governs the behavior of the system. Here, when needed to illustrate
the calculations, we shall use 
\begin{equation}
\label{eq:phi4}
H[\varphi] = \frac{r}{2}\,\varphi^2 + \frac{\lambda}{4!}\,\varphi^4
\end{equation}
which describes a zero-dimensional scalar $\varphi^4$ theory. 

The stochastic force $\eta$ has a Gaussian distribution of zero mean and
second moment fixed by the Einstein's relation
\begin{equation}
\label{eq:einst}
\langle \eta(t)\,\eta(t') \rangle = 2 T\,\delta(t-t')
\end{equation}
where $T$ is the temperature. Indeed, as can be verified by means of the 
associated Fokker-Planck equation \cite{risken},
eq. (\ref{eq:einst}) guarantees that
the time dependent probability
distribution function $P[\varphi,t]$ eventually converge for $t\to\infty$ 
to the equilibrium distribution
\begin{equation}
\label{eq:equil-distr}
P_{\rm eq}[\varphi] \propto \exp(-\beta\,H[\varphi]),
\end{equation}
where $\beta = T^{-1}$.

Given an initial condition $\varphi(t_0)$ the expectation value 
of a generic observable ${\cal O}[\varphi]$ over the stochastic process 
generated by the Langevin equation (\ref{eq:lang}) can be written as
the path integral 
\cite{MarSigRos73,DeDom75,DeDom76,Janssen76,BauJanWag76,DeDomPel78}
\begin{equation}
\label{eq:average}
\langle{\cal O}\rangle = \int \, {\cal D}\varphi\,{\cal D}\hat{\varphi}\
                      {\cal O}[\varphi]\, {\rm e}^{-S[\varphi,\hat{\varphi}]}
\end{equation}
where 
\begin{equation}
S[\varphi,\hat{\varphi}] = \int_{t_0}^{\infty}\, dt\
          \left\{ \hat{\varphi}\left[\partial_t\varphi + 
          \frac{\delta H[\varphi]}{\delta \varphi(t)}\right] 
         - \hat{\varphi}\,T\,\hat{\varphi}\right\}
\end{equation}
The dynamical functional 
$S[\varphi,\hat{\varphi}]$ takes the form of a statistical field theory with
two independent sets of fields, the original field $\varphi$ and the
response field $\hat{\varphi}$, so that all
well established machinery of statistical and quantum field theory
can be applied.
The field $\hat{\varphi}$ is called ``response field'' since it
is conjugated to an external field $h$. As a consequence arbitrary 
response functions can be generated by taking correlators involving 
$\hat{\varphi}$-fields. 
In particular the dynamic susceptibility or response function
$G(t,s)$ reads
\begin{equation}
G(t,s) = \left.\frac{\delta\langle\varphi(t)\rangle}{\delta h(s)}\right|_{h=0}
       = \beta\, \langle\varphi(t)\,\hat{\varphi}(s)\rangle,
\qquad t > s.
\end{equation}
From now on we absorb the factor $\beta$ into the definition of the response 
function.

We note that 
in deriving the dynamical functional $S[\varphi,\hat{\varphi}]$ we have assumed 
the Ito prescription for the stochastic calculus. This implies that
$G(t,t) = 0$. Accordingly, in a perturbative expansion of averages 
(\ref{eq:average}) all
diagrams with at least one loop of the response function can be neglected.

In equilibrium the response function $G(t,s)$ is related to the 
two point correlation function $C(t,s) = \langle\varphi(t)\,\varphi(s)\rangle$
by the fluctuation-dissipation theorem (FDT):
\begin{equation}
\label{eq:FDT}
G(t,s) = \theta(t-s) \frac{\partial}{\partial s}\,C(t,s)
       = -\theta(t-s) \frac{\partial}{\partial t}\,C(t,s)
\end{equation}
We note that 
time translation invariance of equilibrium implies that all two-points
correlators are function of the time difference only. This functional 
dependence is always assumed, even if not explicitly shown, every times 
we consider equilibrium correlators.

In equilibrium the calculation of the two point correlation function 
$C(t,t')$ can be reduced to that of the dynamic self-energy 
$\Sigma_{\hat{\varphi}\hat{\varphi}}(t,t')$. 
Indeed by using standard methods of statistical field theory \cite{zinn} 
and the FDT relation (\ref{eq:FDT}) one obtains for $C(t,t')$
the following equation, valid for $t>t'$:
\begin{equation}
\label{eq:dyna}
\bigl[\partial_t + H''[0] - \Sigma_{\hat{\varphi}\hat{\varphi}}(t,t)\bigr]\, 
            C(t,t') 
 + \int_{t'}^{t}\,ds\, \Sigma_{\hat{\varphi}\hat{\varphi}}(t,s)\,
\partial_s C(s,t') = 0
\qquad t > t' 
\end{equation}
where
\begin{equation} 
H''[0] \equiv \left. 
               \frac{\delta^2 H[\varphi]}{\delta\varphi^2}\right|_{\varphi=0} 
\end{equation}
is the quadratic part of the Hamiltonian $H[\varphi]$. For simplicity we 
assume that in equilibrium $\langle\varphi\rangle = 0$.

In glassy systems the correlation function $C(t,t')$ does not 
vanish in long-time limit $t-t'\to\infty$:
\begin{equation}
\lim_{t-t'\to\infty} C(t,t') = C(\infty) \not= 0
\end{equation}
signaling the breaking of the ergodicity \cite{gotze}.
The value of the ``ergodicity breaking parameter'' $C(\infty)$ is
obtained by taking $t-t'\to\infty$ limit of eq. (\ref{eq:dyna}):
\begin{equation}
\label{eq:bifdy1}
\bigl[H''[0] - \Sigma_{\hat{\varphi}\hat{\varphi}}(0)\bigr]\, C(\infty) +
 \bigl[C(\infty) - C(0)\bigr]\, \Sigma_{\hat{\varphi}\hat{\varphi}}(\infty) = 0
\end{equation}
where 
$\Sigma_{\hat{\varphi}\hat{\varphi}}(\infty)$ is the $t-t'\to\infty$ limit of 
$\Sigma_{\hat{\varphi}\hat{\varphi}}(t,t')$, 
while 
$C(0)=C(t,t)$ and 
$\Sigma_{\hat{\varphi}\hat{\varphi}}(0) = 
 \Sigma_{\hat{\varphi}\hat{\varphi}}(t,t)$ are the equal-time value of the
correlation and self-energy. The latter can be eliminated using the
relation
\begin{equation}
\label{eq:bifdy2}
\bigl[H''[0] - \Sigma_{\hat{\varphi}\hat{\varphi}}(0)\bigr]\, C(0) = 1
\end{equation}
that follows form the $t\to t'^+$ limit of equation (\ref{eq:dyna})
and $\lim_{t\to t'^+} \partial_t C(t,t') = -1$. By combining together
eq. (\ref{eq:bifdy1}) and eq. (\ref{eq:bifdy2}) we finally obtain
the bifurcation equation 
\begin{equation}
\label{eq:bifurc}
\frac{C(\infty)}{C(0)} - 
    C(0)\, \Sigma_{\hat{\varphi}\hat{\varphi}}(\infty) 
            \left[1 - \frac{C(\infty)}{C(0)}\right] = 0.
\end{equation}
In this approach the glass transition is signaled by the appearance of 
a non-trivial solution of the bifurcation equation.

The calculation of the dynamic self-energy 
$\Sigma_{\hat{\varphi}\hat{\varphi}}(t,t')$ is in general a non trivial
task and approximations are usually required to deal with the diagrams of the
dynamical perturbation theory. In this paper
we shall show that in the limit of $t-t'\to\infty$ the calculation of 
the dynamic diagrams simplifies and the bifurcation equation (\ref{eq:bifurc})
can be obtained from a purely static calculation of a $n$-times replicated
system with a 1RSB structure.

The paper is organized as follows.
In Section \ref{sec:dynself} we shall study the structure of the equilibrium
diagrammatic expansion of the dynamic self-energy 
$\Sigma_{\hat{\varphi}\hat{\varphi}}(t,t')$. We shall show that for $t>t'$
$\Sigma_{\hat{\varphi}\hat{\varphi}}(t,t')$ can be decomposed into the 
sum of classes of diagrams obtained by a suitable rearrangement and partial
summation of dynamical diagrams. 
In Section \ref{sec:tt-limit} we shall consider the limit $t-t'\to\infty$ of 
the self-energy $\Sigma_{\hat{\varphi}\hat{\varphi}}(t,t')$. Here,
using both a sum rule approach and a diagrammatic approach, we shall prove
that in the limit $t-t'\to\infty$ all time integrals of the diagrammatic 
expansion can be evaluated and that the self-energy 
$\Sigma_{\hat{\varphi}\hat{\varphi}}(\infty)$ is a function of $C(0)$ and 
$C(\infty)$ only.
In Section \ref{sec:replica} we develop the replica calculation. 
We shall show that 
$\Sigma_{\hat{\varphi}\hat{\varphi}}(\infty)$ evaluated in 
Section \ref{sec:tt-limit} is equal to the static self-energy of an
$n$-times replicated system with a 1RSB structure when the limit $n\to 1$ 
is taken.
This result allows us to derive the bifurcation equation (\ref{eq:bifurc}) on
a purely static calculation.

\section{Equilibrium Dynamic Self-Energy: $t-t'$ finite}
\label{sec:dynself}
The self-energy gives the correction to the free theory correlators when 
interactions are taken into account. Its calculation is
in general rather difficult and one is obliged to use some approximation
schemes. 
Following standard field theory procedures \cite{zinn,lebellac}
a systematic perturbative calculation of the self-energy 
can be established in terms of the so called proper vertex functions.
Diagrammatically these quantities are represented by the 
sets of one-particle irreducible (1PI) diagrams,
i.e., by diagrams that
do not split into two subdiagrams by cutting any single propagator line,
with all the external 
incoming $\varphi$ and outgoing $\hat{\varphi}$ legs amputated.

\subsection{Equilibrium Dynamic Diagrams}
\label{sec:dynrule}
The self-energy $\Sigma_{\hat{\varphi}\hat{\varphi}}(t,t')$ is related to the
proper vertex with the two external outgoing $\hat{\varphi}$ legs removed.
The equilibrium diagrammatic expansion of the dynamic self-energy 
$\Sigma_{\hat{\varphi}\hat{\varphi}}(t,t')$ takes then the following form
\begin{equation}
\label{eq:selfen}
\Sigma_{\hat{\varphi}\hat{\varphi}}(t,t') = \
\includegraphics[scale=1.0, bb= 152 603 325 634]{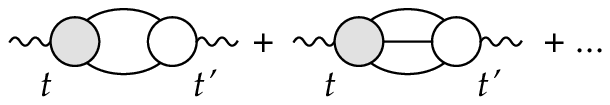}
\end{equation}
where each line connecting the two vertices represents 
the full correlation function \cite{DeDom63,CorJacTom74,Haymaker91}
\begin{equation}
C(s,s') \equiv \langle \varphi(s)\,\varphi(s')\rangle =\ %
\includegraphics[bb= 201 620 263 644]{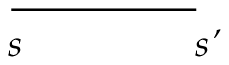}
\end{equation}
The full (left) vertex with $1+r$ external legs 
is the sum of all connected 1PI 
dynamic diagrams with the outgoing $\hat{\varphi}$ leg at $t$,
the wiggly line in (\ref{eq:selfen}), and 
$r$ incoming $\varphi$ legs at times 
$s_1,\ldots, s_r$ ($\leq t$) removed.
To each internal $\varphi\varphi$-line connecting the internal vertices at 
times $s'$ and $s$ is associated the full correlation function 
$C(s,s')$
while to each internal $\varphi\hat{\varphi}$-line
connecting the internal vertices at times $s'$ and $s$ 
is associated the full response function \cite{DeDom63,CorJacTom74,Haymaker91}
\begin{equation}
G(s,s') \equiv \langle \varphi(s)\,\hat{\varphi}(s')\rangle =\ %
\includegraphics[bb= 201 620 263 644]{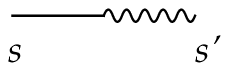}
\end{equation}
with $s > s'$.
All internal times are integrated from $t'$ to $t$.

The empty (right) vertex with $1+r$ external legs is built from the same
1PI diagrams of the full vertex but with all external times 
times $t,s_1\ldots,s_r$ equal to $t'$. 
As a consequence, since in equilibrium the correlation and response functions 
are related by FDT, it is given by the topological equivalent vertex obtained 
from the associated 
static equilibrium theory described by the canonical distribution
(\ref{eq:equil-distr}) \cite{Naketal83}.

The structure just described can be understood as follows. 
First of all we note that since closed loops of response function
vanish, each internal vertex of the 1PI diagrams contributing to the
self-energy $\Sigma_{\hat{\varphi}\hat{\varphi}}(t,t')$ is connected via
$\varphi\hat{\varphi}$-lines to one or the other of the external wiggly lines,
but not to both. This means that the diagram can be divided into two 
subdiagrams made of all the vertices connected to the same external wiggly line.
The two subdiagrams are clearly 1PI and joined together by $2$ or more 
$\varphi\varphi$-lines. 
It is easy to recognize the two subdiagrams as the full/empty vertices.

Alternatively one can invoke the general diagrammatic expansion of the
correlation function $\langle \varphi(t)\,\varphi(t')\rangle$ as the matching 
of two tree expansions \cite{ma},
one for $\varphi(t)$ and one for $\varphi(t')$, 
and note that the lines connecting the full/empty vertices are the lines
joining the two tree expansions.
We note that the assumption that the system is at equilibrium implies 
that all two-times quantities depend only on time difference. Then
the contribution 
from the tree expansion of $\varphi(t')$ averaged over noise and equilibrium 
initial conditions, the empty vertices, cannot depend on time $t'$ and must 
be equal to that obtained from the equilibrium static theory described by the 
canonical distribution (\ref{eq:equil-distr}) with all lines equal
to the static equilibrium correlation function $C(t',t')\equiv C(0)$.

The internal structure of the empty vertices can be inferred
by using the following dynamical functional, see Appendix \ref{app:dynft},
\begin{equation}
\label{eq:eqini}
S'[\varphi,\hat{\varphi}] = S[\varphi,\hat{\varphi}] 
                      + \int_{t_0}^{\infty}\, ds\, H[\varphi(s)]\,\delta(s-t_0)
\end{equation}
to impose statistical equilibrium with the 
canonical distribution (\ref{eq:equil-distr}) at the initial time $t_0$ .
The analysis of the dynamic diagrams generated by $S'[\varphi,\hat{\varphi}]$
shows that 
the effect of the last term is that of canceling out from all dynamic
diagrams for $\Sigma_{\hat{\varphi}\hat{\varphi}}(t,t')$ 
the contribution from times $t_0< s < t'$ yielding
for $\Sigma_{\hat{\varphi}\hat{\varphi}}(t,t')$ the
equilibrium dynamic diagrammatic expansion discussed above.  
This can be understood on a general ground as follows.
The canonical distribution (\ref{eq:equil-distr}) is a stationary 
solution of the associated Fokker-Planck equation, and hence 
the probability distribution of $\varphi$ remains canonical
for any time past $t_0$. Consequence of this is that all quantities evaluated 
from $S'[\varphi,\hat{\varphi}]$ cannot depend on $t_0$.
This guarantees, for example, that all two-times quantities depend 
only on time differences. 
In evaluating $\Sigma_{\hat{\varphi}\hat{\varphi}}(t,t')$ for $t>t'$ we can 
then choose for $t_0$ in (\ref{eq:eqini}) any value $\leq t'$. 
The invariance property ensures that we always get the same diagrams.
Clearly the simplest choice is $t_0=t'$ which, in turn, implies that 
in the diagrammatic dynamical perturbative expansion all free times 
must be integrated from $t'$.
We have seen that the dynamical diagrams for 
$\Sigma_{\hat{\varphi}\hat{\varphi}}(t,t')$ generated by
$S[\varphi,\hat{\varphi}]$
can be divided into two subdiagrams, joined by $r\geq 2$ correlation lines, 
by grouping together all vertices connected  by response 
$\varphi\hat{\varphi}$-lines to the same 
external (amputated) wiggly line at $t$ or $t'$, respectively. 
The dynamical diagrams generated by 
$S'[\varphi,\hat{\varphi}]$ can be divided in a similar way into two
subdiagrams connected by $r$ correlation lines just grouping together 
all vertices connected  by response  $\varphi\hat{\varphi}$-lines to the 
external (amputated) wiggly line at $t$. 
It is easy to realize that this procedure leads to the same
full vertex  obtained from $S[\varphi,\hat{\varphi}]$, 
while the (putative) empty vertex, the one connected to external (amputated) 
leg  at $t'$, contains now only contributions from the last term in 
(\ref{eq:eqini}) since the equal time response function vanishes.
Stated in a different way, the empty vertex contains only
the diagrams generated by the equilibrium initial condition, i.e., it is
equal to topological equivalent 1PI diagram with $1+r$ external (amputated) legs 
of the associated static theory described by the canonical distribution
(\ref{eq:equil-distr}) with all $\varphi\varphi$-lines equal
to the static equilibrium correlation function $C(t,t) \equiv C(0)$.
The same conclusion can be obtained by first dividing the dynamical diagrams
generated by $S'[\varphi,\hat{\varphi}]$ as described above, and then
taking the limit $t_0\to t'$ directly on diagrams.

The definition of empty vertex given above uses $S[\varphi,\hat{\varphi}]$
and follows from the observation that the equilibrium FDT relation between
response and correlation function guarantees that  
setting all external times of a dynamic diagram equal to each other reduces the
dynamic diagram to the topological equivalent diagram of the associated
static equilibrium theory \cite{Naketal83}. 

We can then summarize the rules for writing down the equilibrium dynamic 
diagrams for $\Sigma_{\hat{\varphi}\hat{\varphi}}(t,t')$:

\begin{enumerate}
\item 
Write down the dynamic diagrams generated by the 
      dynamical functional $S[\varphi,\hat{\varphi}]$ 
      using the standard dynamical rules, neglecting all numerical
      symmetry factors.

\item 
In each diagram remove the minimal number of $C$-lines needed to divide 
      the diagram into two disjoint sub-diagrams so that in the first 
      (left) sub-diagram all vertices are connected to time $t$ through 
      $G$-lines, while in the second (right) sub-diagram are connected 
      through $G$-lines to time $t'$.

\item
     In the second (right) sub-diagram, the one connected to $t'$, 
     replace all response $G$-lines by correlation $C$-lines,
     and set all time variables to $t'$.

\item 
     If after replacement two or more different dynamic diagrams lead to 
     the same diagram count the latter only once.

\item
     Multiply each diagram so obtained by the appropriate numerical symmetry 
     factor and evaluate it with the usual rules integrating all left 
     internal times from $t'$ to $t$.  
\end{enumerate}
 
\begin{figure}
\includegraphics[scale=1.0]{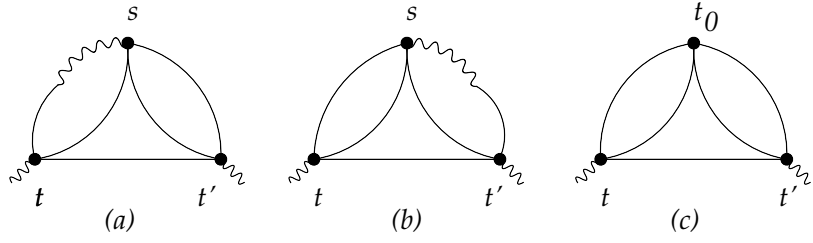}
\caption{Third order dynamic diagrams with equilibrium condition at $t_0$.}
\label{fig:3spic}
\end{figure}

To illustrate the above rules consider the third order dynamic diagrams 
shown in Fig. \ref{fig:3spic} generated by the 
dynamical functional $S'[\varphi,\hat{\varphi}]$ (\ref{eq:eqini})
for the scalar zero-dimensional $\varphi^4$ theory (\ref{eq:phi4}).

By using the standard dynamic rules the contribution of these diagrams is
\begin{eqnarray}
\Sigma_{\hat{\varphi}\hat{\varphi}}^{(3)}(t,t') &=& 
    \frac{1}{2}\int_{t_0}^{t} ds\, G(t,s)\,C(t,s)\,C^2(t',s)\,C(t,t')
    + \frac{1}{2}\int_{t_0}^{t'} ds\, C^2(t,s)\,G(t',s)\,C(t',s)\,C(t,t')
\nonumber\\
&\phantom{+}& 
   + \frac{1}{4} C^2(t,t_0)\,C^2(t',t_0)\,C(t,t')
\end{eqnarray}
With the help  of the FDT relation (\ref{eq:FDT}) the 
contribution can be rewritten as
\begin{eqnarray}
\Sigma_{\hat{\varphi}\hat{\varphi}}^{(3)}(t,t') &=& 
    \frac{1}{4}\int_{t'}^{t} ds\, \partial_s C^2(t,s)\,C^2(t',s)\,C(t,t')
    + \frac{1}{4}\int_{t_0}^{t'} ds\, \partial_s[C^2(t,s)\,C^2(t',s)]\,C(t,t')
\nonumber\\
&\phantom{+}& 
    + \frac{1}{4} C^2(t,t_0)\,C^2(t',t_0)\,C(t,t')
\nonumber\\
&=&
    \frac{1}{2}\int_{t'}^{t} ds\, G(t,s)\,C(t,s)\,C^2(s,t')\,C(t,t')
    + \frac{1}{4} C^3(t,t')\,C^2(t',t')
\label{eq:example1}
\end{eqnarray}

\begin{figure}
\includegraphics[scale=1.0]{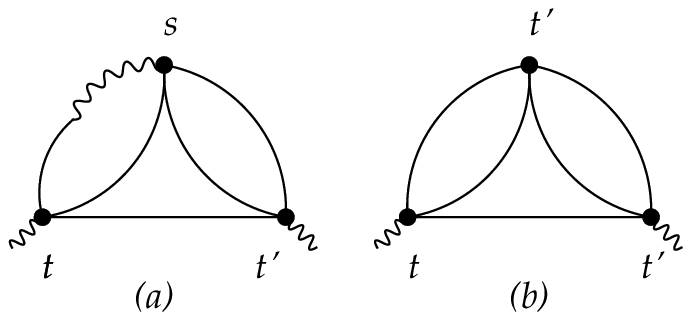}
\caption{Third order equilibrium dynamic diagrams.}
\label{fig:3sped}
\end{figure}

The relevant third order diagrams for the equilibrium dynamic rules, point 1), 
are the first two diagram in Fig. \ref{fig:3spic}.
The decomposition of points 2)-4) leads to the equilibrium 
diagrams shown in Fig. \ref{fig:3sped}. The two diagrams are easily evaluated 
and one recovers the result (\ref{eq:example1}).

\subsection{Self-Energy Base Diagrams}
\label{sec:self-based}
\begin{figure}
\includegraphics[scale=1.0]{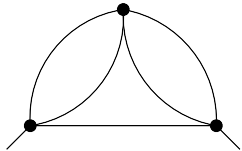}
\caption{Third order base diagram.}
\label{fig:3sbd}
\end{figure}

By reversing the rules to write down the equilibrium dynamic diagrams
it follows that each equilibrium dynamic diagram can be obtained 
by a suitable decoration of a {\sl base diagram}, i.e., of the
diagram with the {\sl same topology} of the
equilibrium dynamic diagram. For example the base diagram for the
dynamic diagrams of the previous example is the one shown in
Fig. \ref{fig:3sbd}.

The equilibrium dynamic diagram is generated from the base diagram by:
\begin{enumerate}

\item 
dividing the base diagram into two sub-diagrams by
cutting some lines to reproduce the topology of point 2) of the
equilibrium dynamic rules.

\item
In the (left) sub-diagram associated to the full vertex attach a
$\hat{\varphi}$-line to each vertex to generate the desired $G$-line 
connection structure. 
\end{enumerate}

It is clear that the base diagram is by construction the dynamic
diagram in which all $G$-lines are replaced by $C$-lines. As a
consequence the self-energy base diagrams are the static self-energy diagrams 
of the associated equilibrium static theory described by the canonical 
distribution (\ref{eq:equil-distr}).

\subsection{Partial summation of full vertex diagrams}
\begin{figure}
\includegraphics[scale=1.0]{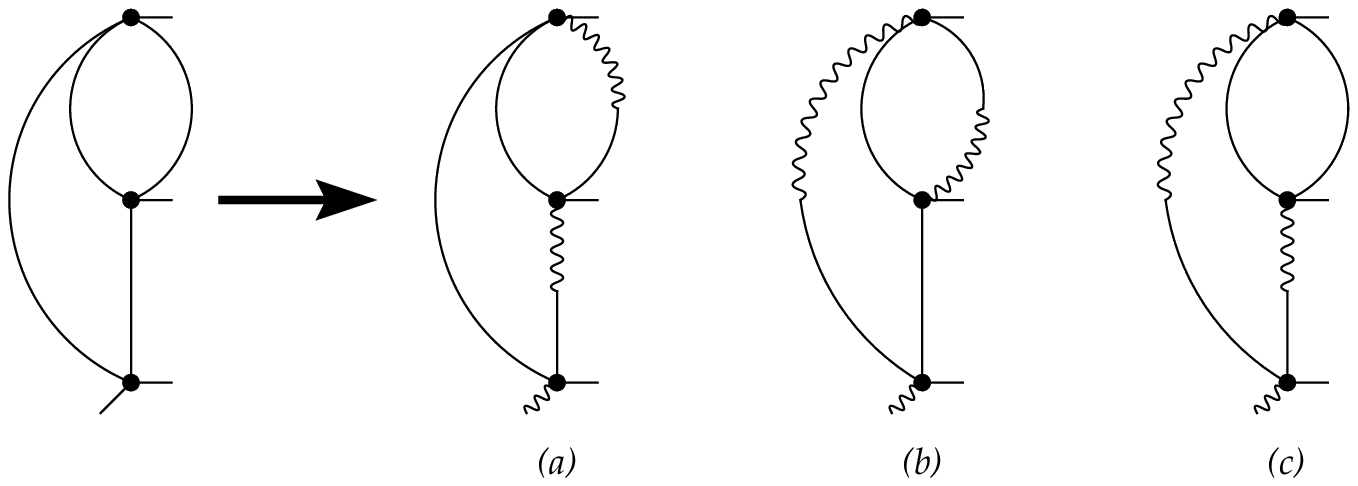}
\caption{A base diagram for the $1+3$ full vertex and the three different
dynamic diagrams that can be generated.
The numerical symmetry factors are not displayed.}
\label{fig:base-vertex}
\end{figure}

From the structure (\ref{eq:selfen}) of the diagrammatic expansion of 
the self-energy it is easy to realize that also
each dynamic diagram contributing to the full
vertex with $1+r$ external legs can be obtained from a suitable (static) 
base diagram by attaching to each internal vertices one 
$\hat{\varphi}$-line to generate the desired $G$-line structure. 
Figure \ref{fig:base-vertex} shows a base diagram for the $1+3$ 
full vertex and the three different dynamic diagrams that can be generated.

A convenient way of imposing that to each vertex is attached one and only one 
$G$-line is by means of anti-commuting Grassman variables.
Following Ref. [\onlinecite{Naketal83}] we then introduce 
for each vertex the pair of conjugated Grassman variables 
$(\alpha_s, \overline{\alpha}_s)$, where $s$ is the time variable
label of the vertex.
By using the FDT relation (\ref{eq:FDT}) 
and the properties of Grassman variables it is easy to see that
\begin{eqnarray}
C[s-s' + \overline{\alpha}_s\alpha_s\theta(s'-s)
      -\overline{\alpha}_{s'}\alpha_{s'}\theta(s-s')] &\phantom{=}&
\nonumber\\
&\phantom{=}& \hskip-3cm
=   C(s-s') 
   + \overline{\alpha}_s\alpha_s\theta(s'-s) \partial_{s'-s} C(s-s')
   + \overline{\alpha}_{s'}\alpha_{s'}\theta(s-s') \partial_{s'-s} C(s-s')
\nonumber\\
&\phantom{=}& \hskip-3cm
= C(s',s) + \overline{\alpha}_s\alpha_s G(s',s) + 
             \overline{\alpha}_{s'}\alpha_{s'} G(s,s').
\label{eq:grass}
\end{eqnarray} 
The last equality follows from even parity of the correlation function 
$C(t,s) = C(s,t)$.
Equation (\ref{eq:grass}) has the following simple diagrammatic representation:
\begin{equation}
\label{eq:grass-dia}
C[s-s' + \overline{\alpha}_s\alpha_s\theta(s'-s)
      -\overline{\alpha}_{s'}\alpha_{s'}\theta(s-s')] =
 \includegraphics[scale=0.8,bb= 201 620 263 644]{corr.eps} %
 + \overline{\alpha}_s\alpha_s 
   \includegraphics[scale=0.8,bb= 201 620 263 644]{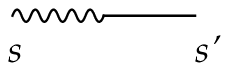}
 + \overline{\alpha}_{s'}\alpha_{s'} 
   \includegraphics[scale=0.8,bb= 201 620 263 644]{resp.eps}
\end{equation}

Consider now a 1PI base diagram $B$ for the $1+r$ full vertex
made of $I$ internal bare vertices, i.e., vertices without external legs, and 
$1+E$ external bare vertices, i.e., vertices with at least one external leg:
\begin{equation}
B := %
\begin{minipage}{0.3\textwidth}
\includegraphics[scale=1.0]{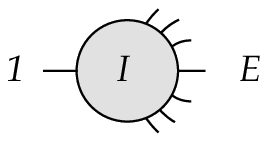}
\end{minipage}
\end{equation}
Each bare external vertex has $n_i=1,2,\ldots$ external legs, with 
$\sum_{i=0}^{E}n_i = 1+r$. Then from (\ref{eq:grass-dia}) and the properties
\begin{equation}
\int d\alpha d\overline{\alpha}\ \overline{\alpha}\alpha = 1,\quad
\int d\alpha\ 1= \int d\overline{\alpha}\ 1 = 0
\end{equation}
of Grassman variables it follows that the sum
$\Lambda_{B}(s_0,...,s_r)$ of all dynamic diagrams 
of the $1+r$ full vertex that can be generated from the base diagram $B$
can be obtained as:

\begin{enumerate}

\item 
  Order the $1+E+I$ bare vertices of $B$ so that: the label $0$ corresponds
  to the external vertex attached to the out-going $\hat{\varphi}$-leg
  at time $t$, the labels $1,\ldots, E$ to the remaining external vertices 
  and labels $E+1,\ldots, E+I$ to the internal vertices.
  
\item
  Assign to vertex $0$ the time variable $s_0=t$, while
  to each vertex labeled by $i=1,\ldots, E+I$ assign the time variable $s_i$
   and the pair of conjugate Grassman variables 
  $(\alpha_{s_i},\overline{\alpha}_{s_i})$.
  From the time-oriented nature of the 
  dynamic diagram it follows that $s_0>s_i>t'$ for $i=1,\ldots, E+I$.

\item
  Assign to each line connecting the vertices $i$ and $j$, with $i,j\not= 0$,
  the function
\begin{equation}
C[s_i-s_j + \overline{\alpha}_{s_i}\alpha_{s_i}\theta(s_j-s_i)
          -\overline{\alpha}_{s_j}\alpha_{s_j}\theta(s_i-s_j)]
\end{equation}
and to each line connected to the vertex $0$ the function
\begin{equation}
C[s_i-s_0 + \overline{\alpha}_{s_i}\alpha_{s_i}\theta(s_0-s_i)]
\end{equation}
with $i,j = 1,\ldots, E+I$.

\item
  Integrate over all Grassman variables,
\begin{equation}
\int\prod_{i=1}^{E+I}\,d\alpha_{s_i} d\overline{\alpha}_{s_i}\, \cdots
\end{equation}
to ensure that all vertices have one, and only one, $\hat{\varphi}$-line
attached to them.

\item
  Integrate all internal time variables $s_i$ 
  from $t'$ to $s_0 = t$,
\begin{equation}
\int_{t'}^{s_0}\prod_{i=E+1}^{E+I}\, d s_i\, \cdots
\end{equation}

\end{enumerate}

When these steps are translated into formulae we end up with:
\begin{eqnarray}
 \Lambda_B(s_0,s_1,\ldots,s_E) &=& M(B)
    \int_{t'}^{s_0}\prod_{i=E+1}^{E+I}\, d s_i\,
    \int\prod_{i=1}^{E+I}\,d\alpha_{s_i} d\overline{\alpha}_{s_i}\,
    \prod_{i=1}^{E+I}\, K_{0i}\, 
 C[s_i-s_0 + \overline{\alpha}_{s_i}\alpha_{s_i}\theta(s_0-s_i)]^{n_{0i}}
\nonumber\\
&\phantom{=}& \hskip1cm\times
\prod_{j=1}^{E+I}\prod_{i=1}^{E+I}\, K_{ji}\,
C[s_i-s_j + \overline{\alpha}_{s_i}\alpha_{s_i}\theta(s_j-s_i)
          -\overline{\alpha}_{s_j}\alpha_{s_j}\theta(s_i-s_j)]^{n_{ji}}
\nonumber \\
&=& M(B)
    \int_{t'}^{s_0}\prod_{i=E+1}^{E+I}\, d s_i\,
    \int\prod_{i=1}^{E+I}\,d\alpha_{s_i} d\overline{\alpha}_{s_i}\,
\nonumber\\
&\phantom{=}& \hskip1.5cm\times
\prod_{j=0}^{E+I}\prod_{i=1}^{E+I}\, K_{ji}\,
C[s_i-s_j + \overline{\alpha}_{s_i}\alpha_{s_i}\theta(s_j-s_i)
          -\overline{\alpha}_{s_j}\alpha_{s_j}\theta(s_i-s_j)]^{n_{ji}}
\label{eq:lambda-b}
\end{eqnarray}
where $M(B)$ is symmetry factor of the base diagram $B$ and
\[
K_{ij} = K_{ji} = \left\{
                \begin{array}{ll}
		  1 & \mbox{if vertices $i$, $j$ are directly connected}\\
		  0 & \mbox{otherwise}
                \end{array}
                          \right.
\]
\[
n_{ij} = n_{ji} = 
        \mbox{\# of lines connecting vertices $i$ and $j$}
\]
give the connection topology of $B$. The last equality in eq. 
(\ref{eq:lambda-b}) follows from the 
observation  that $\theta(s_i-s_0) = 0$ for all $i=1,\ldots, E+I$.

It is easy to verify that the integration over Grassman variables produces
all possible dynamic diagrams, with the correct weighting factor,
that can be generated from the base diagram $B$.

\section{Equilibrium Dynamic Self-Energy: The limit $t-t'\to\infty$}
\label{sec:tt-limit}
\begin{figure}
\includegraphics[scale=0.9]{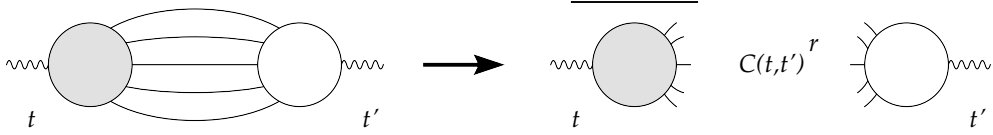}
\caption{Factorization of generic dynamic self-energy diagram for
         $t-t'\to\infty$.}
\label{fig:sigma-dti}
\end{figure}

In the limit $t-t'\to\infty$ some simplifications occur in the
calculation of the dynamic self-energy diagrams. 
Each bare vertex making up the $1+r$ full vertex is connected to the 
bare vertex attached to the external $\hat{\varphi}$-leg at time $t$ through 
$G$-lines
then, since
\begin{equation}
G(s,s') = \theta(s-s')\,\partial_{s'}C(s,s') \not= 0\quad
\mbox{iff $s'\sim s$,}
\end{equation}
it follows that the time variables $s_1,\ldots, s_r$
of the remaining $r$ external legs 
are $s_i\sim t\gg t'$ for $t-t'\to\infty$.
As a consequence in the diagrammatic expansion of the 
self-energy $\Sigma_{\hat{\varphi}\hat{\varphi}}(t,t')$ we can replace 
for $t-t'\to\infty$ all correlation lines connecting the full and 
empty vertex pair with $C(t,t')\equiv C(\infty)$ and the
generic diagram factorizes as shown in Fig. \ref{fig:sigma-dti}.
The overbar on the $1+r$ full vertex means that all external time variables
$s_1,\ldots,s_r$ are integrated from $t'$ to $t$.

The integration in the $1+r$ full vertex can be done considering first 
the contribution of all dynamic diagrams that can be generated from a 
$1+r$ full vertex base diagram $B$ and then summing up the contributions
from all possible base diagrams.
Neglecting the full-empty vertex connection symmetry factor,
the contribution of all dynamic diagram generated by the $1+r$ full vertex 
base diagram $B$ with $1+E$ external and $I$ internal vertices, 
can be written as:
\begin{equation}
\label{eq:limit1}
\int_{t'}^{t}\prod_{i=1}^{E} ds_i\, \Lambda_B(t,s_1,\ldots,s_E)\,
  \prod_{i=1}^{E}\,C(s_i,t')^{n_i}\,   C(t,t')^{n_0-1}\, \Lambda^{(0)}(t')
\end{equation}
where $n_i$ is the number of external legs of the $i$-th external bare
vertex of $B$, and  $\Lambda^{(0)}(t') \equiv\Lambda^{(0)}(C(0))$ 
is the contribution of the $1+r$  empty vertex.\footnote{
        For the empty $1+r$ vertex a decomposition into base
        diagrams similar to that of the full vertex can be done.}

In the limit $t-t'\to\infty$ the first term of eq. (\ref{eq:limit1}) 
is not zero only if all $s_i\sim t \gg t'$ and we can replace all 
$C(s_i,t')$ in the second term with $C(t,t')=C(\infty)$.
Equation (\ref{eq:limit1}) then factorizes, cfr. Fig. \ref{fig:sigma-dti},
as
\begin{equation}
\label{eq:limit2}
\Lambda_B(t)\, C(t,t')^r\, \Lambda^{(0)}(t')
\end{equation}
where
\begin{equation}
\label{eq:lambda-b1}
\Lambda_B(t) = 
\int_{t'}^{t}\prod_{i=1}^{E} ds_i\, 
\left.\Lambda_B(s_0,s_1,\ldots,s_E)\right|_{s_0 = t}.
\end{equation}
Finally by adding the appropriate full-empty vertex connection symmetry
factor and summing eq. (\ref{eq:limit2}) over all possible $1+r$ base 
diagrams $B$ one recovers the complete dynamic diagram contribution 
in the limit $t-t'\to\infty$  from the $1+r$ vertex
to $\Sigma_{\hat{\varphi}\hat{\varphi}}(t,t')$.

\subsection{$1+r$ full vertex base diagram}
From the definition of $\Lambda_B(s_0)$ (\ref{eq:lambda-b1}) and the
expression (\ref{eq:lambda-b}) it follows that
\begin{widetext}
\begin{equation}
 \Lambda_B(s_0) = M(B) 
    \int_{t'}^{s_0}\prod_{i=1}^{N}\, d s_i\,
    \int\prod_{i=1}^{N}\,d\alpha_{s_i} d\overline{\alpha}_{s_i}\,
\prod_{j=0}^{N}\prod_{i=1}^{N}\, K_{ji}\,
C[s_i-s_j + \overline{\alpha}_{s_i}\alpha_{s_i}\theta(s_j-s_i)
          -\overline{\alpha}_{s_j}\alpha_{s_j}\theta(s_i-s_j)]^{n_{ji}}
\label{eq:lambda-b2}
\end{equation}
\end{widetext}
where $N=I+E$ is the total number of bare vertices making up $B$, excluding
the final one connected to the external $\hat{\varphi}$-leg at time $s_0=t$.
This expression can be simplified by integrating over $s_1,\ldots,s_N$ in
a fixed order because from FDT it follows that:
\begin{equation}
\label{eq:grass-2}
C[x-y + \overline{\alpha}_x\alpha_x\theta(y-x)
      - \overline{\alpha}_y\alpha_y\theta(x-y)] = \left\{
  \begin{array}{ll}
C[x-y + \overline{\alpha}_x\alpha_x\theta(y-x)] & \mbox{if $x<y$} \\
\relax &\relax\\
C[y-x + \overline{\alpha}_y\alpha_y\theta(x-y)] & \mbox{if $x>y$} \\
\end{array}\right.
\end{equation} 
%
Thus ordering $s_1,\ldots,s_N$ so that:
\begin{equation}
\label{eq:order}
(s_1,\ldots,s_N) \to
(s_{i_1},\ldots,s_{i_N})\quad :\quad s_{i_p} > s_{i_q} \quad \mbox{if}\ q>p
\end{equation}
the integrand of eq. (\ref{eq:lambda-b2}) can be rewritten as
\begin{eqnarray}
\prod_{j=0}^{N}\prod_{i=1}^{N}\, K_{ji}\, &\phantom{.}&\hskip-.5cm
C[s_i-s_j + \overline{\alpha}_{s_i}\alpha_{s_i}\theta(s_j-s_i)
          -\overline{\alpha}_{s_j}\alpha_{s_j}\theta(s_i-s_j)]^{n_{ji}}
\nonumber\\
&\phantom{.}&
\Rightarrow
\prod_{p=1}^{N}\prod_{q<p}^{0,N}\, K_{i_q i_p}\,
 C[s_{i_p}-s_{i_q} + 
       \overline{\alpha}_{s_{i_p}}\alpha_{s_{i_p}}\theta(s_{i_q}-s_{i_p})
  ]^{n_{i_q i_p}}
\end{eqnarray}
Inserting this expression into (\ref{eq:lambda-b2}) and using the
identity:
\begin{equation}
\int_{t'}^{s_0}\,\prod_{i=1}^Nds_i \equiv 
\sum_{(i_1,\ldots,i_N)\in\mbox{P}(1,\ldots,N)} 
\int_{t'}^{s_0} ds_{i_1} \int_{t'}^{s_{i_1}} ds_{i_2} \cdots 
  \int_{t'}^{s_{i_{N-1}}} ds_{i_N}
\end{equation}
where $\mbox{P}(1,\ldots,N)$ are the $N!$ permutations of $(1,\ldots,N)$,
we end up with
\begin{eqnarray}
 \Lambda_B(s_0) &=& M(B) \sum_{(i_1,\ldots,i_N)\in \mbox{P}(1\ldots,N)}
 \int_{t'}^{s_0} ds_{i_1} \int_{t'}^{s_{i_1}} ds_{i_2} \cdots 
 \int_{t'}^{s_{i_{N-1}}} ds_{i_N}
\nonumber\\
&\phantom{=}& \hskip1.5cm\times
 \prod_{p=1}^{N}\left[\int\,d\alpha_{s_{i_p}} d\overline{\alpha}_{s_{i_p}}\,
\prod_{q<p}^{0,N}\, K_{i_qi_p}\,
 C[s_{i_p}-s_{i_q} + \overline{\alpha}_{s_{i_p}}\alpha_{s_{i_p}}
                  \theta(s_{i_p}-s_{i_q})]^{n_{i_qi_p}}
\right]
\label{eq:lambda-b3}
\end{eqnarray}
The integration over Grassman variables is now diagonal and can
be performed. A straightforward algebra leads to
\begin{eqnarray}
 \Lambda_B(s_0) &=& M(B) \sum_{(i_1,\ldots,i_N)\in \mbox{P}(1\ldots,N)}
 \int_{t'}^{s_0} ds_{i_1} \int_{t'}^{s_{i_1}} ds_{i_2} \cdots 
 \int_{t'}^{s_{i_{N-1}}} ds_{i_N}
\prod_{p=1}^N\, \frac{\partial}{\partial s_{i_p}}\left[
            \prod_{q<p}^{0,N} K_{i_qi_p}\,C(s_{i_q},s_{i_p})^{n_{i_qi_p}}
                      \right]
\nonumber\\
&=& M(B) \sum_{(i_1,\ldots,i_N)\in \mbox{P}(1\ldots,N)}
\prod_{p=1}^N\, \int_{t'}^{s_{i_p-1}} ds_{i_p}\, 
  \frac{\partial}{\partial s_{i_p}}\left[
            \prod_{q=0}^{p-1} K_{i_qi_p}\,C(s_{i_q},s_{i_p})^{n_{i_qi_p}}
                      \right].
\label{eq:lambda-b4}
\end{eqnarray}

\begin{figure}
\includegraphics[scale=1.0]{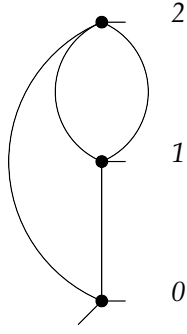}
\caption{Base diagram vertex numbering.}
\label{fig:base-vertex-1}
\end{figure}

As simple example of eq. (\ref{eq:lambda-b4}) consider 
the base diagram $B$ shown in
Fig. \ref{fig:base-vertex}. 
The bare vertices of $B$ are numbered as shown in Fig. \ref{fig:base-vertex-1}.
There are two possible orderings, namely: $\{012\}$ and $\{021\}$.
We use the short-hand notation $0$ for $s_0$, $1$ for $s_1$ and so on.
Then from eq. (\ref{eq:lambda-b4}) the contribution of this base diagram is
\begin{equation}
\Lambda_B(s_0) = \frac{1}{2}\int_{t'}^0d1\int_{t'}^1d2\,
                   \partial_1C(0,1)\,\partial_2\left[C(0,2)C(1,2)^2\right]
+
\frac{1}{2}\int_{t'}^0d2\int_{t'}^2d1\,
                   \partial_2C(0,2)\,\partial_1\left[C(0,1)C(2,1)^2\right]
\label{eq:exampl1}
\end{equation}
The factor $1/2$ is the symmetry factor of the base diagram $B$.

In the direct calculation we have to evaluate the three dynamic diagram 
generated by $B$ shown in Fig. \ref{fig:base-vertex}
using the dynamic rules and FDT,
and then add the results. The first diagram from 
left, with the vertex 
numbered as in Fig. \ref{fig:base-vertex-1}, leads to:
\begin{equation}
  \int_{t'}^0d1\int_{t'}^1d2\, G(0,1)G(1,2)C(1,2)C(0,2) =
 \frac{1}{2}
  \int_{t'}^0d1\int_{t'}^1d2\, \partial_1C(0,1)\partial_2C(1,2)^2C(0,2)
\end{equation}
while the second to:
\begin{equation}
  \int_{t'}^0d2\int_{t'}^2d1\, G(0,2)G(2,1)C(2,1)C(0,1) =
 \frac{1}{2}
  \int_{t'}^0d2\int_{t'}^2d1\, \partial_2C(0,2)\partial_1C(2,1)^2C(0,1)
\end{equation}
and finally the third to:
\begin{equation}
\frac{1}{2}\int_{t'}^0d1\int_{t'}^0d2\, G(0,1)G(0,2)C(2,1)^2 =
 \frac{1}{2}
  \int_{t'}^0d1\int_{t'}^0d2\, \partial_1C(0,1)\partial_2C(0,2)C(2,1)^2
\end{equation}
The factor $1/2$ in the third diagram follows from the symmetry of the 
diagram.
By adding the three contributions, and using the identity
\begin{equation}
\int_{t'}^0d1\int_{t'}^0d2 = 
  \int_{t'}^0d1\int_{t'}^1d2 + \int_{t'}^0d2\int_{t'}^2d1
\end{equation}
one easily recovers the result (\ref{eq:exampl1}).

We note that all integrations in eq. (\ref{eq:exampl1}) can be done. Indeed
the two integrals in eq. (\ref{eq:exampl1}) can be combined together to give
\begin{equation}
\Lambda_B(s_0) = \int_{t'}^0d1\int_{t'}^1d2\,
                   \partial_1C(0,1)\,\partial_2\left[C(0,2)C(1,2)^2\right]
\end{equation}
and performing the integral over $2$ and then over $1$ one ends up with
\begin{equation}
\label{eq:val-expl1}
\Lambda_B(s_0) = \frac{1}{2}\Bigl[  C(0)^4 - C(0)^2\,C(\infty)^2 
               - 2\, C(0)\,C(\infty)^3 + 2\,C(\infty)^4\Bigr]
\end{equation}
where $C(0) = C(t,t)$ and $C(\infty) = C(t,t')$.

The possibility of carrying out all integrals over $s_i$ is not a properties
of this special example, but it is a general result valid for any base diagram 
$B$, as shown in the next subsections.

\subsection{Recursive integration: sum rules approach}
\label{sec:sum-rule}
Equation (\ref{eq:lambda-b4}) can be written as
\begin{equation}
\label{eq:lambda-b5}
\Lambda_B(s_0) = \sum_{(i_1,\ldots,i_N)\in \mbox{P}(1,\ldots,N)}
   F_B\{s_0;s_{i_1},\ldots,s_{i_N}\}
\end{equation}
where
\begin{equation}
\label{eq:effe-b}
F_B\{s_0;s_{i_1},\ldots,s_{i_N}\} = M(B)
\prod_{p=1}^N\, \int_{t'}^{s_{i_p-1}} ds_{i_p}\, 
\frac{\partial}{\partial s_{i_p}}\left[
            \prod_{q=0}^{p-1} K_{i_qi_p}\,C(s_{i_q},s_{i_p})^{n_{i_qi_p}}
                      \right]
\end{equation}
is contribution of the base diagram $B$ when the time variables 
$s_1,\ldots,s_N$ are ordered as $(s_{i_1},\ldots,s_{i_N})$, 
see eq. (\ref{eq:order}).
We shall denote this ordered base diagram by $B\{s_0,s_{i_1},\ldots,s_{i_N}\}$

The sum (\ref{eq:lambda-b5}) can be performed by first
selecting all terms where the pair
$(i_{N-1},i_N)$ is equal to $(h,k)$ or $(k,h)$, and then summing over all
possible distinct index pairs $h,k$.
If we do this we end up with
\begin{equation}
\label{eq:sum2}
\Lambda_B(s_0) = \sum_{(h,k)} 
         \sum_{(i_1,\ldots,i_{N-2})\in \mbox{P}[(1\ldots,N)-(h,k)]}
   \Bigl[
   F_B\{s_0;s_{i_1},\ldots,s_{i_{N-2}}, s_h, s_k\} 
 +
   F_B\{s_0;s_{i_1},\ldots,s_{i_{N-2}}, s_k, s_h\}
\Bigr]
\end{equation}
where the first sum is over all distinct pairs $(h,k)$ with $h,k=1,\ldots,N$,
while the second sun over is all permutations of $(1,\ldots,N)$ once 
$h$ and $k$ are taken out.

From eq. (\ref{eq:effe-b}) it is easy to realize that the two terms in
square brackets differ only for the last two integrals over $s_h$ and $s_k$, 
so that we have to evaluate the sum:
\begin{equation}
\int_{t'}^{s_{i_{N-2}}}ds_h\,
     \partial_{s_h}\left[\prod_{q=0}^{N-2}K_{i_qh}\,C(s_{i_q},s_h)^{n_{i_qh}}
                   \right]
\int_{t'}^{s_h}ds_k\,
     \partial_{s_k}\left[\prod_{q=0}^{N-2}K_{i_qk}\,C(s_{i_q},s_k)^{n_{i_qk}}\,
                                    K_{hk}\,C(s_h,s_k)^{n_{hk}}
                   \right]
+ (h \leftrightarrow k)
\end{equation}
where  the second term is obtained from the first by exchanging 
$s_h$ and $s_k$.

By performing the $s_k$ integral in the first term and the $s_h$ integral
in the second term and rearranging the terms we obtain the
{\it sum rule}
\begin{equation}
\label{eq:sumr2}
F_B\{\ldots,s_h,s_k\} + F_B\{\ldots,s_k,s_h\} =
       F_B\{\ldots,s_h|s_k=s_h\}
     - F_B\{\ldots,s_h|s_k=t'\}
     - F_B\{\ldots,s_k|s_h=t'\}
\end{equation}
where the dots ``$...$'' is the short hand notation for 
``$s_0;s_{i_1},\ldots,s_{i_{N-2}}$''.
The term 
$F_B\{...,s_h|s_k=s_h\}$ is the contribution of the ordered base diagram 
$B\{s_0,s_{i_1},\ldots,s_{i_{N-2}},s_h|s_k=s_h\}$
obtained from $B\{s_0,s_{i_1},\ldots,s_{i_{N-2}},s_h,s_k\}$ by setting 
$s_k=s_h$, i.e., by replacing all lines connecting vertices $h$ and $k$ by 
$C(s_h,s_h)=C(0)$. This reduces the number of integrations by one
since the vertices $h$ and $k$ ``collapse'' into a single composite or
{\it effective} vertex.
Similarly
$F_B\{...,s_h|s_k=t'\}$ is the contribution of the ordered base diagram 
$B\{s_0,s_{i_1},\ldots,s_{i_{N-2}},s_h|s_k=t'\}$ 
obtained from $B\{s_0,s_{i_1},\ldots,s_{i_{N-2}},s_h,s_k\}$ by setting 
$s_k=t'$, i.e., by replacing all lines connecting the vertex $k$ 
to any other vertex $i\not=k$ of $B$ by $C(s_i,t')=C(\infty)$.
Also in this case the number of integrations is reduced by one since
the vertex $k$ is ``removed'' from the diagram.
$F_B\{...,s_k|s_h=t'\}$ is defined in a similar way.

This procedure can be repeated by considering triples $(l,h,k)$ of vertices
and writing
\begin{equation}
\label{eq:lambda-bs3}
\Lambda_B(s_0) = \sum_{(l,h,k)} 
         \sum_{(i_1,\ldots,i_{N-3})\in \mbox{P}[(1\ldots,N)-(l,h,k)]}
  \sum_{\mbox{P}(l,h,k)}\,
  F_B\{s_0;s_{i_1},\ldots,s_{i_{N-3}}, s_l,s_h,s_k\}
\end{equation}
where now the first sum runs over all {\it distinct} triplets $(l,h,k)$,
the second over all possible permutations of $(1,\ldots,N)$ once
the triplet $(l,h,k)$ has been removed and the last over the six
permutations of the triplet $(l,h,k)$.

By using the sum rule (\ref{eq:sumr2}) the last sum can be written as
\begin{eqnarray}
\sum_{\mbox{P}(l,h,k)}\,
F_B\{\ldots, s_l,s_h, s_k\} &=&
\nonumber\\
&\phantom{=}&\phantom{+}\hskip-3cm
F_B\{\ldots,s_l,s_h|s_k=s_h\} - F_B\{\ldots,s_l,s_h|s_k=t'\}
                             - F_B\{\ldots,s_l,s_k|s_h=t'\}
\nonumber\\
&\phantom{=}&\hskip-3cm
+ F_B\{\ldots,s_h,s_k|s_l=s_k\} - F_B\{\ldots,s_h,s_k|s_l=t'\}
                             - F_B\{\ldots,s_h,s_l|s_k=t'\}
\nonumber\\
&\phantom{=}&\hskip-3cm
+ F_B\{\ldots,s_k,s_l|s_h=s_l\} - F_B\{\ldots,s_k,s_l|s_h=t'\}
                             - F_B\{\ldots,s_k,s_h|s_l=t'\}
\label{eq:sum3}
\end{eqnarray}

The  three term with positive sign can be combined together and give
\begin{widetext}
\begin{eqnarray}
F_B\{\ldots,s_l,s_h|s_k=s_h\} + 
F_B\{\ldots,s_h,s_k|s_l=s_k\} +
F_B\{\ldots,s_k,s_l|s_h=s_l\} &=& \nonumber\\
&\phantom{=}&\hskip-9cm \phantom{-}
F_B\{\ldots,s_l|s_h=s_l,s_k=s_h\}
 \nonumber\\
&\phantom{=}&\hskip-9cm
 - F_B\{\ldots,s_l|s_h=t',s_k=s_h\}
 - F_B\{\ldots,s_h|s_k=t',s_l=s_k\}
 - F_B\{\ldots,s_k|s_l=t',s_h=s_l\}
\end{eqnarray}
\end{widetext}
The term $F_B\{\ldots,s_l|s_h=s_l,s_k=s_h\}$ is the contribution of the
ordered diagram $B\{\ldots,s_l|s_h=s_l,s_k=s_h\}$ obtained from
$B\{\ldots,s_l,s_h|s_k=s_h\}$ by setting $s_h=s_l$, or equivalently
from $B\{\ldots,s_l,s_h,s_k\}$ by setting $s_k=s_h$ and $s_h=s_l$.
In both cases the number of integrations is reduced by two and all
lines connecting the three vertices are replaced by $C(0)$.

The term $F_B\{\ldots,s_l|s_h=t',s_k=s_h\}$ is the contribution of the
ordered diagram $B\{\ldots,s_l|s_h=t',s_k=s_h\}$ obtained from
$B\{\ldots,s_l,s_h|s_k=s_h\}$ by setting $s_h=t'$, or equivalently
from $B\{\ldots,s_l,s_h,s_k\}$ by setting $s_k=s_h$ {\it first} and 
{\it then} $s_h=t'$. This means that all lines connecting the
vertices $h$ and $k$ are replaced by $C(0)$ while all
lines connecting either vertex $h$ or $k$ with any other
vertex $i\not= h,k$ by $C(\infty)$, reducing at the same time
the number of integrations by two.
The other two terms are obtained in a similar way.

Finally the six terms in (\ref{eq:sum3}) with negative sign 
can be evaluated in pairs using the sum rule (\ref{eq:sumr2}).
One then gets, for example,
\begin{eqnarray}
F_B\{\ldots,s_l,s_h|s_k=t'\} + F_B\{\ldots,s_h,s_l|s_k=t'\} &=& 
\nonumber\\
&\phantom{=}&\hskip-5.5cm \phantom{-}
F_B\{\ldots,s_l|s_h=s_l,s_k=t'\}
     - F_B\{\ldots,s_l|s_h=t',s_k=t'\}
     - F_B\{\ldots,s_h|s_l=t',s_k=t'\}
\end{eqnarray}
Even if the notation should be now clear 
it is useful to stress that $F_B\{\ldots,s_l|s_h=s_l,s_k=t'\}$ is
the contribution from the ordered diagram $B\{\ldots,s_l|s_h=s_l,s_k=t'\}$
obtained from $B\{\ldots,s_l,s_h|s_k=t'\}$ by setting $s_h=s_l$, 
or equivalently from $B\{\ldots,s_l,s_h,s_k\}$ by setting 
$s_k=t'$ {\it first} and only {\it then} $s_h=s_l$. 
Similarly $F_B\{\ldots,s_l|s_h=t',s_k=t'\}$ is
the contribution from the ordered diagram $B\{\ldots,s_l|s_h=t',s_k=t'\}$
obtained from $B\{\ldots,s_l,s_h|s_k=t'\}$ by setting $s_h=t'$, 
or equivalently from $B\{\ldots,s_l,s_h,s_k\}$ by setting 
$s_k=t'$ and only {\it then} $s_h=t'$. 
The order in the construction is relevant since, for example,
$F_B\{\ldots,s_l|s_h=t',s_k=t'\}$ is in general
different from $F_B\{\ldots,s_l|s_h=t',s_k=s_h\}$.

Collecting all terms we end up with the sum rule:
\begin{widetext}
\begin{eqnarray}
\sum_{\mbox{P}(l,h,k)}\,
F_B\{\ldots, s_l,s_h, s_k\} &=&   
\nonumber\\
&\phantom{=}&\hskip-2cm\phantom{-}
F_B\{\ldots,s_l|s_h=s_l,s_k=s_h\}
\nonumber\\
&\phantom{=}&\hskip-2cm
- F_B\{\ldots,s_l|s_h=t', s_k=s_h\}
- F_B\{\ldots,s_h|s_k=t', s_l=s_k\}
- F_B\{\ldots,s_k|s_l=t', s_h=s_l\}
\nonumber\\
&\phantom{=}&\hskip-2cm
- F_B\{\ldots,s_l|s_h=s_l,s_k=t'\}
+ F_B\{\ldots,s_l|s_h=t', s_k=t'\}
+ F_B\{\ldots,s_h|s_l=t', s_k=t'\}
\nonumber\\
&\phantom{=}&\hskip-2cm
- F_B\{\ldots,s_h|s_k=s_h,s_l=t'\}
+ F_B\{\ldots,s_h|s_k=t', s_l=t'\}
+ F_B\{\ldots,s_k|s_h=t', s_l=t'\}
\nonumber\\
&\phantom{=}&\hskip-2cm
- F_B\{\ldots,s_k|s_l=s_k,s_h=t'\}
+ F_B\{\ldots,s_k|s_l=t', s_h=t'\}
+ F_B\{\ldots,s_l|s_k=t', s_h=t'\}
\label{eq:sumr3}
\end{eqnarray}
\end{widetext}
that, when inserted into eq. (\ref{eq:lambda-bs3}), eliminates two 
time integrations and replaces the lines connected with the vertex
integrated out by either $C(0)$ or $C(\infty)$.

The procedure can be iterated to build sum rules for four vertices,
five vertices and so. 
Despite the fact that the derivation of the sum rules for any number of
vertices is straightforward we shall not push it here because these 
can be more easily obtained using the diagrammatic rules of next subsection.

By using iteratively the sum rules and
\begin{equation}
\label{eq:sumr1}
F_B\{s_0;s|\ldots\} =
F_B\{s_0|s=t,\ldots\} - F_B\{s_0|s=t',\ldots\}
\end{equation}
all $N$ time integrals in $\Lambda_B(s_0)$ can be eliminated in turn 
in favor of $C(0)$ and $C(\infty)$.
This concludes the proof that for any base diagram $B$ all
time integrals can be performed and moreover 
\begin{equation}
\Lambda_B(s_0) \equiv \Lambda_B(C(0),C(\infty))
\end{equation}
where $\Lambda_B(x,y)$ is a function that depends only on the topology of $B$.

To illustrate the sum rule approach we conclude this subsection 
by reconsidering the simple example 
of Fig. \ref{fig:base-vertex-1}.
For this diagram $N=2$, therefore from eq. (\ref{eq:sum2}) and
the sum rule (\ref{eq:sumr2}) it follows
\begin{eqnarray}
\Lambda_B(s_0) &=& F_B\{0;1,2\} +F_B\{0;2,1\}
\nonumber\\
&=&
   F_B\{0;1|2=1\} 
 - F_B\{0;1|2=t'\} - F_B\{0;2|1=t'\}.
\end{eqnarray}
By using now the sum rule (\ref{eq:sumr1}) we end up with
\begin{eqnarray}
\Lambda_B(s_0) &=& \phantom{-} F_B\{0|1=0,2=1\} - F_B\{0|1=t',2=1\}
\nonumber\\
&\phantom{=}&
  - F_B\{0|1=0,2=t'\} + F_B\{0|1=t',2=t'\}
\nonumber\\
&\phantom{=}&
  - F_B\{0|2=0,1=t'\} + F_B\{0|2=t',1=t'\}.
\nonumber\\
\end{eqnarray}
which for the diagram of Fig. \ref{fig:base-vertex-1} leads to
\begin{eqnarray}
\Lambda_B(s_0) &=& \frac{1}{2}\Bigl[
   C(0)^4 - C(0)^2 C(\infty)^2 - C(\infty)^3 C(0)
 + C(\infty)^4 - C(\infty)^3 C(0) + C(\infty)^4 \Bigr]
\nonumber\\ 
&=& \frac{1}{2}\Bigl[
   C(0)^4 - C(0)^2\, C(\infty)^2 - 2\,C(0)\,C(\infty)^3  + 2\,C(\infty)^4
               \Bigr]
\end{eqnarray}
One easily recognizes the result (\ref{eq:val-expl1}) from the
direct calculation.

\subsection{Recursive integration: diagrammatic approach}
While the integration based on sum rules described in the previous subsection
can be carried on for any base diagram $B$, in practical calculations
it can become quite cumbersome.  Thus it would be desirable to have a simpler
way of proceeding. In general diagrammatic methods are simpler and
more transparent, for this reason in this subsection 
we present the diagrammatic 
approach that allow for a {\it graphical} integration directly on the base 
diagram $B$.

The diagrammatic integration goes through the following steps:

\begin{description}

\item $\bullet$ {\sl Time Ordering}\\
  The first step is to generate all possible ordered base diagrams
  $B\{s_0,s_{i_1},\ldots, s_{i_N}\}$ by considering all possible
  permutations $(i_1,\ldots,i_N)\in \mbox{P}(1,\ldots,N)$.
  Once the time order of $B$ has been fixed, each ordered
  diagram $B\{s_0,s_{i_1},\ldots, s_{i_N}\}$ is decorated by 
  orienting each line connecting two vertices by drawing an arrow in 
  the direction of the time flow, i.e., pointing from the shortest to the 
  largest time.

\end{description}

The expression of
$F_B\{s_0;s_{i_1},\ldots,s_{i_N}\}$ is recovered by
first assigning to each line of the oriented diagram 
$B\{s_0,s_{i_1},\ldots,s_{i_N}\}$ the correlation function 
$C(s_{i_p},s_{i_q})$, where  $i_p$ and $i_q$ are the labels of the two 
vertices connected by the line,
and then taking the product for $p=1,\ldots,N$
of the derivative with respect to $s_{i_p}$ of the product of the 
correlation functions $C(s_{i_q},s_{i_p})$ associated with all outgoing 
oriented lines originating from vertex $i_p$.
Finally $F_B\{s_0;s_{i_1},\ldots,s_{i_N}\}$ follows by integrating the
result over all $s_{i_p}$ starting from $s_{i_N}$
and proceeding towards $s_{i_1}$ in order
[cfr. eq. (\ref{eq:effe-b})].
%
In each diagram the integration starts from the vertex with the shortest time.
By construction this vertex has only outgoing oriented lines,
and there is only one of such vertex in each diagram.
The integration cancels the arrows from all oriented lines
originating from the vertex we are integrating on, and replace
the original diagram by {\it two} new diagrams. Thus we have the
following graphical integration rule.

\begin{description}
\item $\bullet$ {\sl Time Integration}

  \begin{enumerate}
  \item
    In each diagram peak up the vertex with only outgoing oriented lines
    and replace the original diagram by the two diagrams obtained as follows:

    \begin{description}
    \item {\sl Diagram $1$:}
      Assign to the vertex with only outgoing oriented lines the next shortest 
      time, i.e., $s_{i_p}\to s_{i_{p-1}}$.
      If the vertex $i_{p-1}$ is not directly connected to $i_p$
      draw a {\it dotted} line between the two vertices to remember
      they have the same time.
      Next replace all outgoing oriented lines with simple not-oriented lines
      drawing them as: 
      {\it full} line if the line connects two vertices at different time; 
      {\it dashed} line if the line connects two vertices at the same time.
      As before,
      we shall call {\it effective} vertex 
      the group of bare vertices with equal time, i.e., the
      group of vertices connected by dashed or dotted lines.
      
    \item {\sl Diagram $2$:}
      Set the time of the vertex with only outgoing oriented lines to $t'$ and
      replace all outgoing oriented lines of the vertex 
      by cutted lines, i.e.,  replace the arrow by the ``cut'' sign ``$/$''. 
      Finally multiply the diagram by $-1$ so that two new diagrams
      contribute with opposite signs.
    \end{description}
    
  \item 
    Merge all diagrams that differ only for the oriented 
    outgoing lines originating from the same effective vertex into the single
    diagram obtained from any one of them by orienting all
    not-oriented full lines originating from the effective vertex according to 
    the orientation of all diagrams we are merging to. 
    This leaves us with diagrams made of
    only oriented lines (arrows), dashed or dotted lines and cutted
    lines.

  \end{enumerate}
\end{description}
It is not difficult to realize that this rule produces diagrams
where the vertex, or effective vertex, with the shortest time
has only outgoing oriented lines. 
Moreover the vertices connected via cutted lines
do not contribute anymore to the integration process.
The procedure can then be
iterated untill one is left with diagrams made of only  dashed or 
cutted lines. The dotted lines are used for time bookkeeping and
can be eliminated if not needed.

\begin{figure}
\includegraphics[scale=0.7]{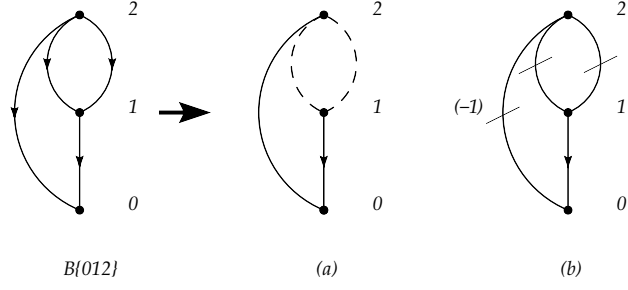}
\caption{Oriented base diagram $B\{012\}$ and result of first diagrammatic
         integration.}
\label{fig:gi-ex1-012}
\end{figure}

\begin{figure}
\includegraphics[scale=0.7]{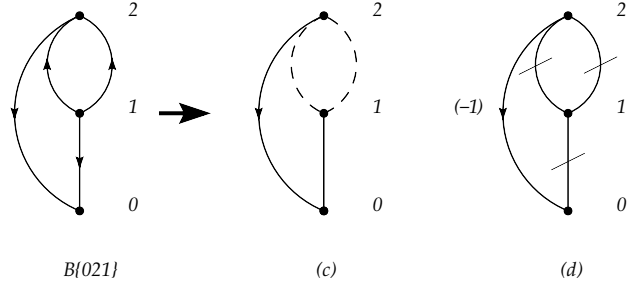}
\caption{Oriented base diagram $B\{021\}$ and result of first diagrammatic
         integration.}
\label{fig:gi-ex1-021}
\end{figure}

\begin{figure}
\includegraphics[scale=0.7]{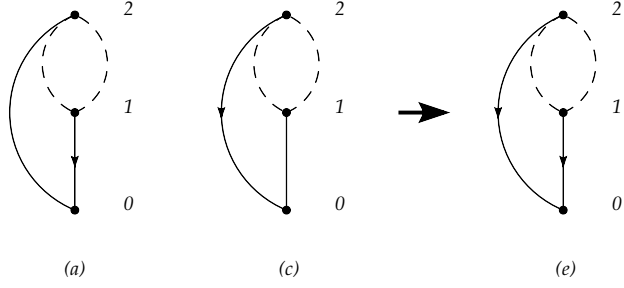}
\caption{Merging of diagram $(a)$ of Fig. \protect\ref{fig:gi-ex1-012} and
         diagram $(c)$ of Fig. \protect\ref{fig:gi-ex1-021}.
        }
\label{fig:gi-ex1-merge}
\end{figure}

\begin{description}
\item $\bullet$ {\sl Value of diagrams}\\
  The value of $\Lambda_B(s_0)$ is obtained by
  evaluating each final diagram with {\it dashed} lines replaced by
  the equal time correlation function $C(t,t) = C(0)$ and {\it cutted} lines
  by the infinite time correlation function $C(t,t')=C(\infty)$, and summing up
  all contributions from different diagrams with the appropriate 
  plus or minus sign.
\end{description}
To illustrate the graphical integration rules we consider again the base 
diagram shown in Fig. \ref{fig:base-vertex-1}. The two possible oriented 
diagrams $B\{012\}$ and $B\{021\}$ are shown in Figs. \ref{fig:gi-ex1-012} and
\ref{fig:gi-ex1-021} together with the result of step $1)$
of the time integration rule.
The diagrams $(a)$ in Fig. \ref{fig:gi-ex1-012} and $(c)$ in 
Fig. \ref{fig:gi-ex1-021} differs only for the orientation of the
outgoing lines originating from the effective vertex made by the bare
vertices $1$ and $2$. The two diagrams are then merged together 
into the diagram shown in Fig. \ref{fig:gi-ex1-merge}, step $2)$ of the
time integration rule.

\begin{figure}
\includegraphics[scale=0.7]{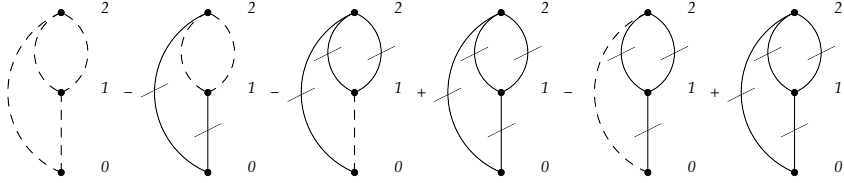}
\caption{Final diagrams.
        }
\label{fig:gi-ex1-tot}
\end{figure}

Diagrams $(b)$ of Fig. \ref{fig:gi-ex1-012}, $(d)$ of Fig. \ref{fig:gi-ex1-021}
and $(e)$ of Fig. \ref{fig:gi-ex1-merge} contain only (effective) 
vertices with all outgoing lines or connected by cutted lines, 
thus the time integration steps
$1)$ and $2)$ can be repeated for each one of these diagrams.  This second round
eliminates all oriented lines and, taking into account the
$-1$ signs, $\Lambda_B(s_0)$ is given by the sum of the six diagrams shown
in Fig. \ref{fig:gi-ex1-tot}.
Evaluating these diagrams and multiplying the result by the symmetry factor 
$M(B)=1/2$ of the base diagram, we and up with
\begin{eqnarray}
\Lambda_B(s_0) &=& \frac{1}{2}\Bigl[
   C(0)^4 - C(0)^2\,C(\infty)^2 - C(0)\,C(\infty)^3 
 + C(\infty)^4- C(0)\,C(\infty)^3 + C(\infty)^4
 \Bigl]
\nonumber\\
 &=&
      \frac{1}{2}\Bigl[
    C(0)^4 - C(0)^2\,C(\infty) - 2\,C(0)\,C(\infty)^3 + 2\,C(\infty)^4
                 \Bigr],
\end{eqnarray}
i.e., 
again the result (\ref{eq:val-expl1}) obtained previously by direct 
integration.

\section{The replica calculation}
\label{sec:replica}
\subsection{The $1+r$ full vertex base diagram}
In the previous Section we have shown that in the limit $t-t'\to\infty$ 
the contribution $\Lambda_B(t)$ of the $1+r$ base vertex $B$ to the 
$1+r$ full vertex can be obtained by considering all possible diagrams 
that can be constructed from $B$ by replacing the lines connecting the 
bare vertices by either {\it dashed} or {\it cutted} lines in all possible ways.
Clearly there are some constraints to be considered, for example all lines
connecting the same pair of vertices must be replaced 
{\it simultaneously}. In other words it not possible to have dashed {\it and}
cutted line connecting the same pair of vertices,
see e.g. Fig. \ref{fig:gi-ex1-tot}.

Given any base diagram $B$ a simple way of generating all allowed diagrams is 
that of attaching to each bare vertex $i=0,1,\ldots,N$ of $B$ a label $a_i$.
Each label can take the values $1,\ldots,n$, where $n$ is an arbitrary integer.
The next step is that of introducing the ``projector'' operators 
$\delta_{a_ia_j}$ and $(1-\delta_{a_ia_j})$ and associate to each line 
of $B$ connecting the vertex $i$ and vertex $j$ the ``correlation'' function
\begin{equation}
\label{eq:corr-rep}
C_{a_ia_j} = C(0)\,\delta_{a_ia_j} + C(\infty)\,(1-\delta_{a_ia_j})
\end{equation}
By representing the projector $\delta_{a_ia_j}$ with a dashed line and
the projector $(1-\delta_{a_ia_j})$ with a cutted line it is easy to 
convince oneself that the sum over all $a_i$ ($i=1,\ldots,N$) from 
$1$ to $n$ generates all possible diagrams that can be constructed 
from $B$. 

The interesting point of this method is that it does
not only generates all possible diagrams but, as a ``bonus'',
it also gives the correct weight and sign for each one of them.
For a single line this is rather trivial since
for $n\to 0$ one has the formal identity:
\begin{equation}
\label{eq:formal-int}
\int_{t'}^{t}ds\,\frac{\partial}{\partial s}\,C(t,s) = C(0) - C(\infty)
 \equiv \lim_{n\to 0} \sum_{b=1}^{n} C_{ab}
\end{equation}
that ensures that for $n\to 0$ an oriented line is 
not only replaced by a dashed line and a cutted line, but also that
the second one comes with a negative relative sign.
By using the properties of the projection operators
this identity can be extended to a bunch of oriented lines connecting 
two vertices.

In the Appendix \ref{app:limit0} we show that the
sum over labels $a_i$ with $i=1,\ldots,N$ for any base diagram $B$
reproduces in $n\to 0$ limit the sum rules discussed in 
Sec. \ref{sec:sum-rule}. This implies that
$\Lambda_B(s_0)$ can be expressed in the ``replica'' form
\begin{equation}
\label{eq:lambda-b-rep}
\Lambda_B(t) = M(B)\,\lim_{n\to 0}
 \sum_{a_1,\ldots,a_N}^{1,n}\prod_{i=1}^{N}\prod_{j=0}^{i-1}
   K_{ij}\,C_{a_ia_j}^{n_{ij}}
\end{equation}
We then have the following simple rule to evaluate $\Lambda_B(s_0)$:
\begin{enumerate}
\item
 Given a base diagram $B$
 assign to each bare vertex $i=0,\ldots,N$ of $B$ a ``replica'' 
 index $a_i$.
\item
Assign to each line connecting the vertices $i$ and $j$ 
the correlation function $C_{a_ia_j}$  given in eq. (\ref{eq:corr-rep}).
\item
Sum over all replica indexes $a_i$ with $i=1,\ldots,N$ from 
$1$ to $n$ and multiply the result by the diagram symmetry factor
$M(B)$.
\item
Take the limit $n\to 0$ of the result.
\end{enumerate}

To illustrate the procedure, i.e., formula (\ref{eq:lambda-b-rep}),
we consider again the base diagram
shown in Fig. \ref{fig:base-vertex-1}. For this diagram eq. 
(\ref{eq:lambda-b-rep}) reads:
\begin{eqnarray}
\Lambda_B(t) &=& \frac{1}{2}\lim_{n\to 0} \sum_{a_1=1}^{n}\sum_{a_2=1}^n
    C_{a_0a_1}\,C_{a_0a_2}\,C_{a_1a_2}^2
\nonumber\\
&=&
  \frac{1}{2}\lim_{n\to 0}\left[
    C(0)^4 + (n-1) C(0)^2C(\infty)^2 + 2(n-1)C(0)\,C(\infty)^3
     +(n-1)(n-2)C(\infty)^4
\right]
\nonumber\\
&=&
   \frac{1}{2}\left[
    C(0)^4 - C(0)^2C(\infty)^2 - 2C(0)\,C(\infty)^3 + 2C(\infty)^4
\right]
\end{eqnarray}
We note that the sum produces all not-equivalent diagrams
shown of Fig. \ref{fig:gi-ex1-tot}.
Indeed the first term in the second line that follows from $a_2=a_1=a_0$ 
corresponds to the first diagram shown in Fig. \ref{fig:gi-ex1-tot}.
The weight $1$ reflects the fact that there is only one possible choice
$a_2=a_1=a_0$. 
Similarly the second term obtained for $a_2=a_1\not=a_0$
corresponds to the second diagram in Fig. \ref{fig:gi-ex1-tot}.
There are $n-1$ choices that satisfy the constraint $a_2=a_1\not=a_0$. 
The third term follows from either $a_1=a_0\not=a_2$ or
$a_1\not=a_0=a_2$ and indeed corresponds to the third and fifth
diagrams of Fig. \ref{fig:gi-ex1-tot}. In both cases there are $n-1$ choices
that satisfy the constraint.
Finally  the last term is obtained by taking all three indexes $a_0,a_1$ and
$a_2$ different from each others, and hence $(n-1)(n-2)$ possible choices. 
This term corresponds to the fourth (or the equivalent sixth) diagram of Fig. 
\ref{fig:gi-ex1-tot}.

\subsection{The self-energy diagrams}
To finalize the calculation of the self-energy diagrams (\ref{eq:selfen}) 
in the limit $t-t'\to\infty$ we also need the contribution
from the empty vertex, the quantity $\Lambda^{(0)}(t')$ in eq. 
(\ref{eq:limit2}).
The empty vertex is given by the equilibrium diagrams of the associated
static theory described by the canonical distribution (\ref{eq:equil-distr}),
as a consequence the empty vertex is made by the base diagrams used 
to construct the full-vertex with all lines equal to the equal-time 
correlation $C(0)$.\footnote{
  The empty $1+r$ vertex can be obtained by setting all external times
  of the full vertex equal to $t'$. This ensures that the full dynamic
  diagram reduces to its static counterpart, i.e., the empty vertex.
  See, e.g., Ref. [\onlinecite{Naketal83}].}

The contribution from any one of such diagram $B$ to the empty vertex 
can be readily written in the replica formalism:
\begin{equation}
\label{eq:lambda-b0-rep}
\Lambda_{B}^{(0)}(t') = M(B)
    \prod_{i=1}^{N}\prod_{j=0}^{i-1} K_{ij}\,C_{aa}^{n_{ij}}
\end{equation}
where we used the same vertex numbering convention as the full vertex
diagrams. The limit $n\to 0$ is not necessary since we are only using the
diagonal part of $C_{ab}$.

By using the representations (\ref{eq:lambda-b-rep}) and 
(\ref{eq:lambda-b0-rep}) the contribution 
$\Lambda_{B_fB_e}^{(r)}(t,t')$ of the self-energy diagram with the
$1+r$ full vertex generated by the base diagram $B_f$ and the $1+r$ empty 
vertex generated by the base diagram $B_e$ reads in the limit $t-t'\to\infty$,
see eq. (\ref{eq:limit2}),
\begin{equation}
\Lambda_{B_fB_e}^{(r)}(t,t') = 
M(B_f)\,M(B_e)\,M(B_f,B_e)\lim_{n\to 0}
 \sum_{a_1,\ldots,a_N}^{1,n}\prod_{i=1}^{N_f}\prod_{j=0}^{i-1}
   K^f_{ij}\,C_{a_ia_j}^{n^f_{ij}}\,
   C(\infty)^r\,
    \prod_{i=1}^{N_e}\prod_{j=0}^{i-1} K^{e}_{ij}\,C_{bb}^{n^e_{ij}}
\end{equation}
where $M(B_f,B_e)$ is the symmetry factor generated by the connections between
the two vertices.

The term $C(\infty)^r$ follows from the lines connecting the $1+E_f$ external
vertices of $B_f$ to the $1+E_e$ external vertices of $B_e$. 
Thus by assuming that
the replica indexes $a_i$ of $B_f$ never take a value equal to that of 
the replica index $b$ of $B_e$ this term can be written as
\begin{equation}
 C(\infty)^r \Rightarrow \prod_{i=0}^{E_f}\prod_{j=0}^{E_e}
                         J_{ij}\,C_{a_ib}^{m_{ij}}
\end{equation}
where the rectangular symmetric matrices $J_{ij}$ and $m_{ij}$ give the
topology of the connections between $B_f$ and $B_e$, i.e., 
$J_{ij} = 1$ if the two external vertices are connected or $0$ otherwise,
while $m_{ij}=1,2,\ldots$ ($\sum_{ij}m_{ij} = r$) gives the number of 
lines between the two vertices.

The values of the replica indexes of the two vertices can be made different 
either assuming $b=n+1$ or forcing the two sets of replicas to assume
different values by adding the projectors $(1-\delta_{a_ib})$.
This second method is more appealing since it maintains the symmetry 
of replica index, the little price to pay is that now the replica index of 
diagram $B_f$ can take only $n-1$ values, so the limit must be changed
from $n\to 0$ to $n\to 1$.

Collection all terms we end up with
\begin{eqnarray}
\label{eq:lambda-bb-rep}
\Lambda_{B_fB_e}^{(r)}(t,t') &=& 
M(B_f)\,M(B_e)\,M(B_f,B_e)\lim_{n\to 1}
 \sum_{a_1,\ldots,a_N}^{1,n}\left[\prod_{i=1}^{N_f}\prod_{j=0}^{i-1}
   K^f_{ij}\,C_{a_ia_j}^{n^f_{ij}}\right]\,
\left[\prod_{i=0}^{E_f}(1-\delta_{a_ib}) \prod_{j=0}^{E_e}
                         J_{ij}\,C_{a_ib}^{m_{ij}}\right]
\nonumber\\
&\phantom{=}&\phantom{=======================}\times
   \left[ \prod_{i=1}^{N_e}\prod_{j=0}^{i-1} K^e_{ij}\,C_{bb}^{n^e_{ij}}\right]
\end{eqnarray}

In Sect. \ref{sec:self-based} we have seen that the equilibrium dynamic diagrams
for $\Sigma_{\hat{\varphi}\hat{\varphi}}(t,t')$ can be obtained by dividing 
a suitable base diagram with the same topology of the equilibrium dynamic 
diagrams into two sub-diagrams, one leading to $B_f$ and the other to $B_e$.
This is precisely the role of the projector
$\prod_{i=0}^{E_f}(1-\delta_{a_ib})$ in eq. (\ref{eq:lambda-bb-rep}).
The total contribution from the self-energy base diagram $B$ 
is now obtained by considering all possible divisions of $B$ into $B_f$
and  $B_e$ and summing up the result.
It is not difficult to realize that all possible distributions of 
the bare vertices of $B$ between $B_f$ and $B_e$ can be generated by 
considering all possible insertions of the projector
$\prod_{i=0}^{E_f}(1-\delta_{a_ib})$.
As a consequence the total contribution 
$\Sigma_{\hat{\varphi}\hat{\varphi}}^{(B)}(t,t')$ 
to the self-energy from all equilibrium diagrams generated by the base 
diagram $B$ in the limit $t-t'\to\infty$ reads
\begin{equation}
\label{eq:sigma-final}
\Sigma_{\hat{\varphi}\hat{\varphi}}^{(B)}(t,t') = M(B) \lim_{n\to 1}
\sum_{a_1,\ldots,a_N} \prod_{i=1}^{N+1}\prod_{j=0}^{i-1}
   K_{ij}\,C_{a_ia_j}^{n_{ij}}\, \left(1-\delta_{a_0a_{N+1}}\right)
\end{equation}
where we have associated the index $0$ with the external vertex at
$t$ and the index $N+1$ to the external vertex at $t'$ so that 
the total number of bare vertices of $B$ is $N+2$.

The formula \label{eq:final} has a simple meaning: 
to obtain the contribution from
all diagrams generated by the self-energy base diagram $B$ just
attach to each vertex of $B$ a replica index $a$ and to each line
connecting the vertex $a$ and $b$ the correlation function $C_{ab}$ 
[eq. (\ref{eq:corr-rep})].
Then sum over all replica indexes
from $1$ to $n$ keeping the replica index of the vertices attached
to the $t$ and $t'$ external legs fixed and different from each other.
At the end take the limit $n\to 1$.

To illustrate the procedure we consider the base diagram $B$ shown
in Fig. \ref{fig:3sbd}. Equation (\ref{eq:sigma-final}), with the
symmetry factor $M(B)=1/4$, gives
\begin{eqnarray}
 \Sigma_{\hat{\varphi}\hat{\varphi}}^{(3)}(t,t') &=& 
\frac{1}{4}\lim_{n\to 1}
  \sum_{a_1=1}^{n} C_{a_0a_1}^2\,C_{a_0a_2}\,C_{a_1a_2}^2\,(1-\delta_{a_0a_2})
\nonumber\\
&=& \frac{1}{4}\lim_{n\to 1}
 \left[ 2\,C(0)^2\,C(\infty)^3 + (n-2)\,C(\infty)^5 \right]
\nonumber\\
&=& \frac{1}{4}
 \left[ 2\,C(0)^2\,C(\infty)^3 - \,C(\infty)^5 \right].
\end{eqnarray}
The base diagram of Fig. \ref{fig:3sbd} produces the equilibrium dynamic 
diagrams
shown in Fig. \ref{fig:3sped} whose value is given by eq. (\ref{eq:example1}).
Evaluating the integral for $t-t'\to \infty$ we indeed have
\begin{eqnarray}
 \Sigma_{\hat{\varphi}\hat{\varphi}}^{(3)}(t,t') &=& 
 \frac{1}{2}\int_{t'}^{t}ds\, \frac{\partial}{\partial s}C(t,s)\ C(t,s)
            C(\infty)^3 
  +\frac{1}{4}C(0)^2\,C(\infty)^3
\nonumber\\
&=&
 \frac{1}{4}\left[C(0)^2 - C(\infty)^2\right]\,C(\infty)^3 
  +\frac{1}{4}C(0)^2\,C(\infty)^3
\nonumber\\
&=& \frac{1}{4}
 \left[ 2\,C(0)^2\,C(\infty)^3 - \,C(\infty)^5 \right].
\end{eqnarray}

\begin{figure}
\includegraphics[scale=1.0]{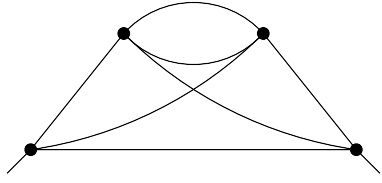}
\caption{Fourth order crossed self-energy base diagram.}
\label{fig:cross}
\end{figure}

As further example consider the base diagram of Fig. \ref{fig:cross}.
This diagram has a symmetry factor of $1/2$, thus the replica calculation
gives
\begin{eqnarray}
\label{eq:cross4c}
 \Sigma_{\hat{\varphi}\hat{\varphi}}^{(4c)}(t,t') &=& 
\frac{1}{2}\lim_{n\to 1}
  \sum_{a_1=1}^{n}\sum_{a_2=1}^{n} 
     C_{a_0a_1}\,C_{a_0a_2}\,C_{a_1a_2}^2\,C_{a_0a_3}\,C_{a_1a_3}\,C_{a_2a_3}
      (1-\delta_{a_0a_3})
\nonumber\\
&=&
\frac{1}{2}\lim_{n\to 1} \left[
  2\,C(0)^4\,C(\infty)^3 + n\,C(0)^2\,C(\infty)^5 + 4(n-2)\,C(0)\,C(\infty)^6
  + (n-2)(n-3)\,C(\infty)^7
                         \right]
\nonumber\\
&=&
\frac{1}{2}\left[
  2\,C(0)^4\,C(\infty)^3 + C(0)^2\,C(\infty)^5 - 4\,C(0)\,C(\infty)^6
  + 2\,C(\infty)^7
                         \right]    
\end{eqnarray}
The direct calculation of the equilibrium dynamic diagrams generated by this
base diagram is shown in Appendix \ref{app:cross}.

\subsection{The replicated system and the bifurcation equation}
The self-energy base
diagram $B$ are the {\it static} self-energy diagrams generated 
by the Hamiltonian $H$ of the associated equilibrium static theory described 
by the canonical distribution (\ref{eq:equil-distr}). 
Thus from the results discussed so far it follows that 
\begin{equation}
\label{eq:st-dyn}
\lim_{t-t'\to\infty} \Sigma_{\hat{\varphi}\hat{\varphi}}(t,t') =
\lim_{n\to 1} \Sigma_{ab}, \qquad a\not=b
\end{equation}
where $\Sigma_{ab}$ is the static self-energy 
of the $n$ times replicated system described by the Hamiltonian 
${\cal H}_n = \sum_{a=1}^nH[\varphi_{a}]$ 
and the 1RSB correlation function [eq. (\ref{eq:corr-rep})]
\begin{equation}
\label{eq:c-rep}
C_{ab} = \langle\varphi_a\varphi_b\rangle_{\rm eq}
       = C(0)\,\delta_{ab} + C(\infty)\,(1-\delta_{ab})
\end{equation}
with $C(0)=C(t,t)$ the equal time value of the equilibrium two time 
correlation function $C(t,t')$ and
$C(\infty)$ its $t-t'\to\infty$ limit.
The average $\langle\cdots\rangle_{\rm eq}$ is taken with the canonical 
probability distribution function
\begin{equation}
P_{\rm eq}^{n}\propto \exp\left(-\beta\,{\cal H}_n\right).
\end{equation}

It is now straightforward to derive the bifurcation equation. 
The correlation $C_{ab}$ and the self-energy $\Sigma_{ab}$ are related 
by the Dyson equation
\begin{equation}
\label{eq:dyson}
\sum_{c}\,\left[[C^{0}]_{ac}^{-1} - \Sigma_{ac}\right]\,C_{cb} = \delta_{ab}
\end{equation}
where $[C^{0}]_{ab}^{-1} = H''(0)\delta_{ab}$.
The self-energy $\Sigma_{ab}$, given by the sum of all 1PI diagrams
generated by the vertices of ${\cal H}_n$ and correlation $C_{ab}$, has 
the 1RSB structure:
\begin{equation}
\label{eq:s-rep}
\Sigma_{ab} = \Sigma_1\,\delta_{ab} 
            + \Sigma_0\,(1-\delta_{ab}).
\end{equation}
where the diagonal part $\Sigma_1 = \Sigma_{aa}$ is equal to the static 
self-energy of a single replica, i.e., to the static self-energy 
$\Sigma_{\hat{\varphi}\hat{\varphi}}(0)$ of the original system. 
By inserting eqs. (\ref{eq:c-rep}) and (\ref{eq:s-rep}) into the Dyson 
equation (\ref{eq:dyson}), and summing over the replica index $c$ with 
$a\not=b$ we have
\begin{equation}
\left[H''(0) - \Sigma_1\right]\,C(\infty)
 - \Sigma_0\, C(0)
 - (n-2)\,\Sigma_0\, C(\infty)
 = 0
\end{equation}
which, using (\ref{eq:st-dyn}), in the $n\to 1$ limit reduces to 
\begin{equation}
\label{eq:dyaneqb}
\left[H''(0) - \Sigma_{\hat{\varphi}\hat{\varphi}}(0)\right]\,C(\infty)
 +\left[C(\infty) - C(0)\right]\,
      \Sigma_{\hat{\varphi}\hat{\varphi}}(\infty) = 0.
\end{equation}
Setting now $a=b$ in the Dyson equation (\ref{eq:dyson}) and taking the limit 
$n\to 1$ we end up with
\begin{equation}
\label{eq:dyaeqb}
\left[H''(0) - \Sigma_{\hat{\varphi}\hat{\varphi}}(0)\right]\,C(0)
 - (n-1)\Sigma_0 C(\infty) \stackrel{n\to 1}{=}
\left[H''(0) - \Sigma_{\hat{\varphi}\hat{\varphi}}(0)\right]\,C(0) = 1.
\end{equation}
that with eq. (\ref{eq:dyaneqb}) leads back to the 
bifurcation equation (\ref{eq:bifurc}) derived from dynamics.

\section{Conclusions}
In this paper motivated by the replica formalism developed for the analysis
of glassy system, with or without quenched disorder, we have shown that
the long time limit $C(\infty) = \lim_{t-t'\to\infty} C(t,t')$ 
of the equilibrium two time correlation function $C(t,t')$ can be computed
from the $n\to 1$ limit of the static of a $n$-times replicated system with a
1RSB structure. In particular we have shown that the Dyson equation
of the replicated system leads in the $n\to 1$ limit to the bifurcation
equation for the ergodicity breaking parameter $C(\infty)$ derived from
dynamics.
The main result is the equivalence of the $t-t'\to\infty$ limit
of the equilibrium dynamic self-energy 
$\Sigma_{\hat{\varphi}\hat{\varphi}}(t,t')$ with the
$n\to 1$ limit of the off-diagonal static self-energy $\Sigma_{ab}$ of a
$n$-times replicated system with a 1RSB structure.

The proof is based on the analysis of $t-t'\to\infty$ limit of
the self-energy diagrams of the dynamic field theory perturbation 
approach, and follows the following steps.
We first discussed the general structure of the equilibrium dynamical 
perturbation diagrammatic expansion of the self-energy 
$\Sigma_{\hat{\varphi}\hat{\varphi}}(t,t')$ and shown that this can
be written in terms of groups of dynamic diagrams 
obtained by a common base diagram. Next we derived an explicit expression 
valid for any $t$ and $t'$ for the contribution of each group.
In the limit $t-t'\to\infty$ all time integrals can be evaluated and
for each group one obtains an expression that depends only on $C(0)$ and 
$C(\infty)$.
We have proved this result by a sum-rule approach and then by a
graphical diagrammatic integration method. 
Next we have shown that the diagrams that result from the diagrammatic 
integration method can be reproduced by introducing ``replicas''.
Moreover by choosing for the replicated system a 1RSB scheme
we are able to reproduce the correct value of each diagram by taking
limit $n\to 1$ for the number $n$ of replicas.
This is proved by showing that in this limit the replica approach leads to the
same sum-rules derived from dynamics.

In this paper we focused on the general proof of the 
equivalence between the replica and the dynamical approaches.
For this reason we did not discussed any physical consequences of
this equivalence. These will be considered in a forthcoming paper
\cite{BirCriprep}.

Finally we observe that the results reported in this paper 
follow from our studies on the Mode Coupling Theory approach to the glass 
transition.
This is why we focused on the self-energy 
$\Sigma_{\hat{\varphi}\hat{\varphi}}(t,t')$. However the method
developed here can be extended to deal with the $t-t'\to\infty$ limit of other 
quantities.

\acknowledgments
The results reported in this paper originate from several discussions I 
had with G. Biroli and C. De Dominicis. I wish to thank both of them for 
the time spent on discussing this subject, and G. Biroli for a critical 
reading of the manuscript.
I also  acknowledge the SPTH of CEA, where part of this work was done, 
for the warm hospitality and support.

\appendix
\section{The limit $n\to 0$}
\label{app:limit0}
In this Appendix we show that the sum over all
replica indices $a_i$ with $i=1,\ldots, N$ of any $1+r$ base diagram $B$ 
reproduces in the limit $n\to 0$ 
the sum rules discussed in Sec. \ref{sec:sum-rule}.

Let us denote by
\begin{equation}
 F_B^{(2)}\{s_h,s_k\} =
   \sum_{(i_1,\ldots,i_{N-2})\in\mbox{P}[(1,\ldots,N)-(h,k)]}
   F_B\{s_0;s_{i_1},\ldots,s_{i_{N-2}},s_h,s_k\}
\end{equation}
the contribution to $\Lambda_B(s_0)$ [eq. (\ref{eq:lambda-b5})]
from all time ordering of
$s_1,\ldots,s_N$ with $s_k$ the smallest time and $s_k$ the next smallest one,
i.e., $s_i > s_h > s_k$ for any $i\not= h,k$.
Then $\Lambda_B(s_0)$ can be written as, see eqs. (\ref{eq:sum2}) and
(\ref{eq:sumr2}),
\begin{eqnarray}
\label{eq:app-a2}
\Lambda_B(s_0) &=& \sum_{(h,k)} \left[
    F_B^{(2)}\{s_h,s_k\} +  F_B^{(2)}\{s_k,s_h\}
                                \right]   
\nonumber\\
&=& \sum_{(h,k)} \left[
    F_B^{(2)}\{s_h|s_k=s_h\} -  F_B^{(2)}\{s_h|s_k=t'\}  
                                 -  F_B^{(2)}\{s_k|s_h=t'\}  
                                \right]   
\end{eqnarray}
We may select in a similar way from the replica sum in eq. 
(\ref{eq:lambda-b-rep})
all terms in which the replica indices $a_h$ and $a_k$ have a value
smaller or equal to all others replica indices $a_i$, $i\not=h,k$.
Then by denoting their sum by:
\begin{equation}
 F^{(2)}_{a_ha_k} = M(B) \left[\prod_{i\not=h,k}^{1,N}\sum_{a_i=1}^{n}\right]
              \prod_{i=1}^{N}\prod_{j=0}^{i-1}\,K_{ij}\,C_{a_ia_j}^{n_{ij}},
\qquad a_{h,k}\leq a_i,\, i\not=h,k
\end{equation}
the equation (\ref{eq:lambda-b-rep}) can be written as:
\begin{equation}
 \Lambda_B(t) = \lim_{n\to 0} \sum_{(h,k)} {\sum_{a_ha_k}}' 
               F^{(2)}_{a_ha_k}
\end{equation}
where the first sum runs over all distinct pairs $h,k$ while the second
over $a_h$ and $a_k$. The prime $'$ over the sum sign 
means that only values of $a_{k,h}\leq a_i$ with $i\not= h,k$ are included
into the sum. Finally by decomposing the 
sums over $a_h$ and $a_k$ as
\begin{equation}
{\sum_{a_ha_k}}' \equiv \sump{a_h=a_k}{-3pt} + \sump{a_h>a_k}{-3pt} 
                       + \sump{a_k>a_h}{-3pt}
\end{equation}
we end up with
\begin{equation}
 \Lambda_B(t) = \lim_{n\to 0} \sum_{(h,k)} \left[
        \sump{a_h=a_k}{-3pt}  F^{(2)}_{a_ha_k}
      + \sump{a_h>a_k}{-3pt}  F^{(2)}_{a_ha_k}
      + \sump{a_k>a_h}{-3pt}  F^{(2)}_{a_ha_k} \right]
\end{equation}
It is easy to realize that the three terms correspond to the three terms
in the second line of eq. (\ref{eq:app-a2}), in the same order. 
The first one is immediate. In the second term $a_k$ is always smaller than
$a_h$ and hence smaller than all $a_i$ with $i\not=k$. This is exactly
the same structure of the second term of the sum rule (\ref{eq:app-a2}). 
Similarly the third one corresponds to the third and last term in 
(\ref{eq:app-a2}). 
Only the signs are different. The correct signs, and hence the correct 
second order sum rule, are recovered in the limit $n\to 0$ since 
in this limit  due to the restriction in the sums 
the second and third terms acquire a negative sign from the 
lines connecting the vertices $h$ and $k$.

In a similar way one recovers the third order sum rule. Indeed
with a straightforward extension of the notation, the third order sum rule, 
eqs. (\ref{eq:lambda-bs3}) and (\ref{eq:sumr3}), can be written as
\begin{eqnarray}
\label{eq:app-a7}
\Lambda_B(s_0) &=& \sum_{(l,h,k)} \sum_{\mbox{P}(l,h,k)}\,
   F_B^{(3)}\{s_0;s_l,s_h,s_k\}
\nonumber\\
&=& \sum_{(l,h,k)} \Bigl[
    F_B^{(3)}\{s_0;s_l|s_h=s_l,s_k=s_h\}
\nonumber\\
&\phantom{=}&\phantom{===}
- F_B^{(3)}\{s_l|s_h=t', s_k=s_h\}
- F_B^{(3)}\{s_h|s_k=t', s_l=s_k\}
- F_B^{(3)}\{s_k|s_l=t', s_h=s_l\}
\nonumber\\
&\phantom{=}&\phantom{===}
- F_B^{(3)}\{s_l|s_h=s_l,s_k=t'\}
+ F_B^{(3)}\{s_l|s_h=t', s_k=t'\}
+ F_B^{(3)}\{s_h|s_l=t', s_k=t'\}
\nonumber\\
&\phantom{=}&\phantom{===}
- F_B^{(3)}\{s_h|s_k=s_h,s_l=t'\}
+ F_B^{(3)}\{s_h|s_k=t', s_l=t'\}
+ F_B^{(3)}\{s_k|s_h=t', s_l=t'\}
\nonumber\\
&\phantom{=}&\phantom{===}
- F_B^{(3)}\{s_k|s_l=s_k,s_h=t'\}
+ F_B^{(3)}\{s_k|s_l=t', s_h=t'\}
+ F_B^{(3)}\{s_l|s_k=t', s_h=t'\}
\Bigr]
\end{eqnarray}
where
\begin{equation}
 F_B^{(3)}\{s_l,s_h,s_k\} =
   \sum_{(i_1,\ldots,i_{N-3})\in\mbox{P}[(1,\ldots,N)-(l,h,k)]}
   F_B\{s_0;s_{i_1},\ldots,s_{i_{N-3}},s_l,s_h,s_k\}
\end{equation}
On the other hand by selecting three replica indexes, the replica
expression (\ref{eq:lambda-b-rep}) leads to
\begin{equation}
 \Lambda_B(t) = \lim_{n\to 0} \sum_{(l,h,k)} \sump{a_la_ha_k}{-4pt}
               F^{(3)}_{a_la_ha_k}
\end{equation}
where 
\begin{equation}
 F^{(3)}_{a_la_ha_k} = M(B) 
             \left[\prod_{i\not=l,h,k}^{1,N}\sum_{a_i=1}^{n}\right]
              \prod_{i=1}^{N}\prod_{j=0}^{i-1}\,K_{ij}\,C_{a_ia_j}^{n_{ij}},
\qquad a_{l,h,k}\leq a_i,\, i\not=l,h,k
\end{equation}
Now by decomposing the sum over the three replica indexes as
\begin{eqnarray}
\sump{a_la_ha_k}{-4pt} &\equiv&
 \sump{a_l=a_h=a_k}{-10pt}
\nonumber\\
&\phantom{\equiv}&
+ \sump{a_l>a_h=a_k}{-10pt} + \sump{a_h>a_k=a_l}{-10pt} + \sump{a_k>a_l=a_h}{-10pt} 
\nonumber\\
&\phantom{\equiv}&
 + \sump{a_l=a_h>a_k}{-10pt} + \sump{a_l>a_h>a_k}{-10pt} + \sump{a_h>a_l>a_k}{-10pt}
\nonumber\\
&\phantom{\equiv}&
+ \sump{a_h=a_k>a_l}{-10pt} + \sump{a_h>a_k>a_l}{-10pt} + \sump{a_k>a_h>a_l}{-10pt} 
\nonumber\\
&\phantom{\equiv}&
 + \sump{a_k=a_l>a_h}{-10pt} + \sump{a_k>a_l>a_h}{-10pt} + \sump{a_l>a_k>a_h}{-10pt}
\end{eqnarray}
one identifies all the terms of the third order sum rule (\ref{eq:app-a7}).
The third order sum rule (\ref{eq:app-a7}) is 
recovered in the limit $n\to 0$ since each inequality in the sum
gives in this limit to a ``minus'' sign which, when combined together,
reproduces the correct sings.

From these two examples it should be clear that the correspondence between the
sum rules derived with the two formulations, dynamic with integrals 
and replica with sum over integers, can be extended to sum rules of any order.
Indeed the multiple sums in eqs. (\ref{eq:lambda-b5}) and 
(\ref{eq:lambda-b-rep}) can be always decomposed in the same way by peaking up
the same group of vertices. 
Moreover the lower limit of the integrals 
in the dynamical formulation is always associated
with an inequality in the correspondent sum of the replica formulation.
Then, since in the limit $n\to 0$ each inequality gives a ``minus'' sign, 
it is clear that in the $n\to 0$ limit both formulations lead to the same
sum rules and hence
to the same expression
for $\Lambda_B(t)$, once the two point correlation function $C_{ab}$ of the
replicated system is identified with (\ref{eq:corr-rep}).

\section{The fourth order crossed diagrams}
\label{app:cross}

In this Appendix we sketch the calculation of the equilibrium dynamic diagrams
generated by the fourth order self-energy base diagram of Fig. \ref{fig:cross}
in the limit $t-t'\to\infty$. The equilibrium dynamic diagrams 
generated by the base diagram of Fig. \ref{fig:cross}
are shown in Fig. \ref{fig:cross-dy}, see Section \ref{sec:dynrule}, and lead 
to:

\begin{figure}
\includegraphics[scale=1.0]{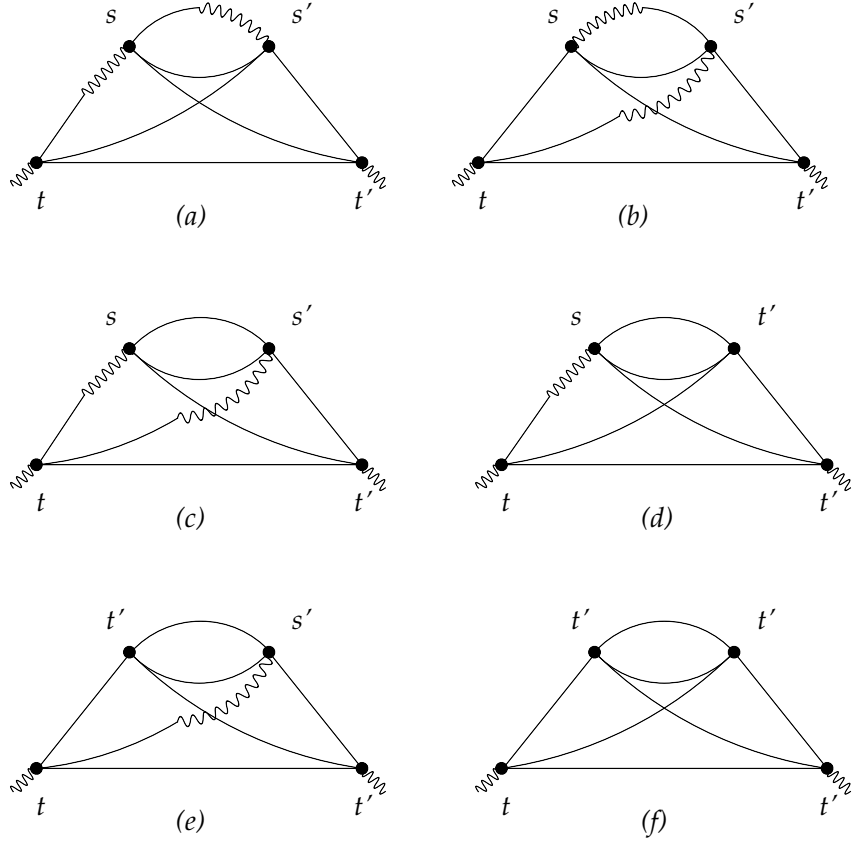}
\caption{Fourth order crossed equilibrium dynamic diagrams.}
\label{fig:cross-dy}
\end{figure}

\begin{eqnarray}
 \Sigma_{\hat{\varphi}\hat{\varphi}}^{(4c)}(t,t') &=& \phantom{+}
 \int_{t'}^{t}ds\int_{t'}^{s}ds'\, 
     G(t,s) C(t,s') G(s,s') C(s,s') C(s,t') C(s',t') C(t,t')
\nonumber\\
&\phantom{=}& +
 \int_{t'}^{t}ds'\int_{t'}^{s'}ds\, 
     C(t,s) G(t,s') G(s',s) C(s',s) C(s,t') C(s',t') C(t,t')
\nonumber\\
&\phantom{=}& + \frac{1}{2}
 \int_{t'}^{t}ds\int_{t'}^{t}ds'\, 
     G(t,s) G(t,s') C(s,s')^2 C(s,t') C(s',t') C(t,t')
\nonumber\\
&\phantom{=}& + \frac{1}{2}
 \int_{t'}^{t}ds\, G(t,s) C(s,t')^3 C(t,t')^2 C(t',t')
\nonumber\\
&\phantom{=}& + \frac{1}{2}
 \int_{t'}^{t}ds'\, G(t,s') C(s',t')^3 C(t,t')^2 C(t',t')
\nonumber\\
&\phantom{=}& + \frac{1}{2}
  C(t,t')^3\,C(t',t')^4.
\end{eqnarray}
The factors $1/2$ are the symmetry factor of the diagram.

The third integral can be split in two by using the identity
\begin{equation}
\int_{t'}^{t}\,ds\,\int_{t'}^{t}\,ds' = 
       \int_{t'}^{t}\,ds\,\int_{t'}^{s}\,ds' + 
\int_{t'}^{t}\,ds'\,\int_{t'}^{s'}\,ds
\end{equation}
By using now the FDT relation $G(s,s') = \partial_{s'}C(s,s')$ the first and 
half of the third integral can be written as:
\begin{eqnarray}
&\phantom{=}& \phantom{x}
 \frac{1}{2} 
 \int_{t'}^{t}ds\int_{t'}^{s}ds'\, 
     \partial_s C(t,s) C(t,s') \partial_{s'} C(s,s')^2 C(s,t') C(s',t') C(t,t')
\nonumber\\
&\phantom{=}& +
 \frac{1}{2}
 \int_{t'}^{t}ds\int_{t'}^{s}ds'\, 
     \partial_s C(t,s) \partial_{s'} C(t,s') C(s,s')^2 C(s,t') C(s',t') C(t,t')
\nonumber\\
&\phantom{=}&\phantom{=====} = 
 \frac{1}{2} 
 \int_{t'}^{t}ds\int_{t'}^{s}ds'\, 
   \partial_s C(t,s) \partial_{s'} [C(t,s') C(s,s')^2] C(s,t') C(s',t') C(t,t')
\end{eqnarray}
In the limit $t-t'\to\infty$ we can replace each of the last three correlation
functions by $C(\infty)$ since the integrand is different from zero only if
$s\sim s'\sim t \gg t'$. The integral over $s'$ can now be easily done
and one gets
\begin{eqnarray}
\label{eq:ac1}
&\phantom{=}&
\frac{1}{2} C(0)^2 C(\infty)^3 \int_{t'}^{t}ds\, \partial_s C(t,s) C(t,s) 
 - \frac{1}{2}C(\infty)^6 \int_{t'}^{t}ds\, \partial_s C(t,s)
\nonumber\\
&\phantom{=}&\phantom{=====} = 
\frac{1}{2} C(0)^2 C(\infty)^3 \left[C(0)^2 - C(\infty)^2\right]
 - \frac{1}{2}C(\infty)^6 \left[C(0) - C(\infty)\right]
\end{eqnarray}

The other half of the third integral can be combined with the second integral,
the one from diagram $(b)$, and a similar manipulation leads again to 
the result (\ref{eq:ac1}). This is not unexpected since diagram $(a)$ and 
$(b)$ can be changed one into the other by exchanging $s$ and $s'$.

The fourth and fifth integrals, i.e., those from diagrams $(d)$ and $(e)$, 
can be readily done with the help of the FDT relation and in both cases
one obtains for $t-t'\to\infty$:
\begin{equation}
\frac{1}{2}C(0)C(\infty)^5 \left[C(0) - C(\infty)\right].
\end{equation}
Collecting all contributions we ends up with 
\begin{equation}
\lim_{t-t'\to\infty} \Sigma_{\hat{\varphi}\hat{\varphi}}^{(4c)}(t,t') = 
\frac{1}{2}\left[
  2\,C(0)^4\,C(\infty)^3 + C(0)^2\,C(\infty)^5 - 4\,C(0)\,C(\infty)^6
  + 2\,C(\infty)^7
                         \right]    
\end{equation}
as found from the replica calculation, see eq. (\ref{eq:cross4c}).

\section{Derivation of eq. (\ref{eq:dyna})}
\label{app:dynft}
Our aim is to study the correlations associated with the stochastic process
(\ref{eq:lang}). Instead of working directly with eq. (\ref{eq:lang}) it is 
more convenient to construct a generating functional from which correlations
can be obtained. Following the standard procedure of field theory \cite{zinn} 
one introduces an external time-dependent source  $J(t)$ and defines
the generating functional\footnote{%
          The temperature $T$ is absorbed into the definition of 
          the response function.}
\begin{equation}
  Z[J] = {\cal N}\,\int {\cal D}\varphi\,{\cal D}\eta\, {\cal P}[\varphi_0]\,
          \delta(\varphi - \varphi_{\eta})\, 
          \exp\left[ - \int_{t_0}^{\infty}\, {\rm d} t\, J\varphi\right]
          \exp\left[ - \int_{t_0}^{\infty}\, {\rm d} t\, \frac{\eta^2}{2} 
              \right]
\label{eq:gf1}
\end{equation}
where $\varphi_{\eta}$ is the solution of stochastic the eq. (\ref{eq:lang})
for a given realization of the stochastic field $\eta(t)$ and 
initial condition $\varphi_0 = \varphi(t_0)$.
The initial condition is assigned with the probability ${\cal P}[\varphi_0]$, 
that we take equal to the equilibrium probability distribution 
(\ref{eq:equil-distr}). Finally ${\cal N}$ is a
normalizing constant. 
The $\delta$-function stands for
\begin{equation}
\label{eq:delta}
  \delta(\varphi - \varphi_{\eta}) = 
         \delta\left[ \frac{\partial\varphi}{\partial t}
                                + \frac{\delta H[\varphi]}{\delta \varphi(t)}
                                       - \eta
                                    \right]\,
      \det\left|\frac{\delta\eta}{\delta\varphi}\right|.
\end{equation}
where the factor
$\det|\delta \eta/\delta\varphi|$ is the Jacobian of the transformation
$\eta\to\varphi$, that in the Ito calculus is equal to one.

By using the integral representation of the delta function with the help 
of the auxiliary hat-field $\hat{\varphi}$, and performing the integral over 
the stochastic field $\eta$, a straightforward algebra leads to
\begin{equation}
\label{eq:genfunc}
  Z[J]= {\cal N} \int \, {\cal D}\varphi\,{\cal D}\hat{\varphi}\
       {\rm e}^{-S'[\varphi,\hat{\varphi}] - 
                 \int_{t_0}^{\infty}\, {\rm d} t\, J\varphi}
\end{equation}
where $S'[\varphi,\hat{\varphi}]$ is given by eq. (\ref{eq:eqini}) and includes
the contribution from the initial equilibrium distribution 
(\ref{eq:equil-distr}). By adding a second external time-dependent source
$\hat{J}(t)$ coupled to $\hat{\varphi}$ all correlations and responses can be
obtained from differentiation of the generating functional
\begin{equation}
\label{eq:genfuncd}
  Z[J, \hat{J}]= {\cal N} \int \, {\cal D}\varphi\,{\cal D}\hat{\varphi}\
       {\rm e}^{-S'[\varphi,\hat{\varphi}] - 
       \int_{t_0}^{\infty}\, {\rm d} t\, [J\varphi + \hat{J}\hat{\varphi}]}
\end{equation}
Define now the $2\times 2$ correlation matrix
\begin{equation}
{\cal G}(t,t') \equiv \left(
                 \begin{array}{cc}
 {\cal G}_{\varphi\varphi}(t,t')  &  {\cal G}_{\varphi\hat{\varphi}}(t,t') \\
  \phantom{=} & \phantom{=} \\
 {\cal G}_{\hat{\varphi}\varphi}(t,t')  &  {\cal G}_{\hat{\varphi}\hat{\varphi}}(t,t') \\
                  \end{array}
                 \right)
                 = \left(
                 \begin{array}{cc}
  \langle\varphi(t)\varphi(t')\rangle &  
  \langle\varphi(t)\hat{\varphi}(t')\rangle \\
  \phantom{=} & \phantom{=} \\
  \langle\hat{\varphi}(t)\varphi(t')\rangle &  
  \langle\hat{\varphi}(t)\hat{\varphi}(t')\rangle \\
                  \end{array}
                 \right)
                 = \left(
                 \begin{array}{cc}
  C(t,t') &  G(t,t') \\
  \phantom{=} & \phantom{=} \\
  G(t',t) & \hat{C}(t,t') \\
                  \end{array}
                 \right).
\end{equation}
The average is taken with the dynamical functional $S'[\varphi,\hat{\varphi}]$.
We assumed for convenience
$\langle\varphi\rangle=\langle\hat{\varphi}\rangle = 0$.
The correlation matrix ${\cal G}(t,t')$ is solution of the Dyson equation
\begin{equation}
\label{eq:dysonab}
\int_{t_0}^{\infty} ds\, \sum_{\gamma}\left[
          G_{\alpha\gamma}^{-1}(t,s) - \Sigma_{\alpha\gamma}(t,s)
                                      \right]\, {\cal G}_{\gamma\beta}(s,t') 
                = \delta_{\alpha\beta}\,\delta(t-t')
\end{equation}
where 
the Greek indexes run over the two values $(\varphi,\hat{\varphi})$, and
the matrix $G_{\alpha\beta}(t,t')$ is the ``free'' correlation matrix 
obtained from the second order term of the expansion of 
$S'[\varphi,\hat{\varphi}]$ 
about $\langle\varphi\rangle=\langle\hat{\varphi}\rangle = 0$.
A simple calculation gives
\begin{equation}
G^{-1}(t,t') \equiv \left(
                 \begin{array}{cc}
 H''[0]\,\delta(t-t_0)\,\delta(t-t') & 
                 \left(-\partial_t + H''[0]\right)\,\delta(t-t') \\
  \phantom{=} & \phantom{=} \\
 \left(\partial_t + H''[0]\right)\,\delta(t-t') & 
       -2 \delta(t-t')
                  \end{array}
                 \right)
\end{equation}
where
$H''[0] \equiv \left. \delta^2 H[\varphi]/\delta\varphi^2\right|_{\varphi=0}$.

The equation for the correlation function $C(t,t')$ follows now from 
the Dyson equation (\ref{eq:dysonab}) by choosing 
$\alpha=\hat{\varphi}$ and $\beta=\varphi$:
\begin{equation}
\label{eq:eqC1}
\bigl[\partial_t + H''[0]\bigr]\, C(t,t') - 2 G(t',t) 
  -\int_{t_0}^{t} ds\, \Sigma_{\hat{\varphi}\varphi}(t,s)\,C(s,t') 
  -\int_{t_0}^{t} ds\, \Sigma_{\hat{\varphi}\hat{\varphi}}(t,s)\,G(t',s) = 0 
\end{equation}
In equilibrium FDT holds, then inserting the FDT relations
\begin{equation}
 G(t,t') = -\theta(t-t')\,\partial_t C(t,t'), \qquad 
 \Sigma_{\hat{\varphi}\varphi}(t,t') = -\theta(t-t')\,
              \partial_t \Sigma_{\hat{\varphi}\hat{\varphi}}(t,t')
\end{equation}
into eq. (\ref{eq:eqC1}) and integrating by parts a straightforward algebra
leads to eq. (\ref{eq:dyna}) of the main text.



\begin{thebibliography}{99}
\bibitem{metastable}
  See for example 
  M. M\'ezard, G. Parisi and M. Virasoro, 
  Spin Glass Theory and Beyond, 
  (World Scientific, Singapore 1987);
  A.P. Young (ed),
  Spin Glasses and Random Fields,
  (World Scientific, Singapore 1998);
  M. Rub\'i, C. Perez-Vicente (eds)
  Complex Behaviour in Glassy Systems,
  (Springer-Verlag, Berlin 1996);
  C.A. Angell, Science {\bf 267} (1995) 1924.

\bibitem{KirThi87}
  T.R. Kirkpatrick and D. Thirumalai,
  Phys. Rev. Lett. {\bf 58}, (1987) 2091.

\bibitem{CriSom92}
  A. Crisanti, and H.J. Sommers, 
  Z.. f\"ur Phys. B {\bf 87}, (1992) 341.

\bibitem{CriHorSom93}
  A. Crisanti, H. Horner, and H.J. Sommers, 
  Z.. f\"ur Phys. B {\bf 92}, (1883) 257.

\bibitem{BouMez94}
  J.P. Bouchaud and M. M\'ezard,
  J. Phys. I (France) {\bf 4}, (1994) 1109.

\bibitem{MarParRit94}
  E. Marinari, G. Parisi and F. Ritort,
  J. Phys. A {\bf 27}, (1994) 7615;
             {\bf 27}, (1994) 7647.

\bibitem{CriSom95}
  A. Crisanti and H.J. Sommers
  J. Phys. I (France) {\bf 5}, (1995) 805.

\bibitem{BouCugKurMez96}
  J.P. Bouchaud, L. Cugliandolo, J. Kurchan and M. M\'ezard,
  Physica A {\bf 226}, (1996) 243.

\bibitem{Monasson95}
  R. Monasson,
  Phys. Rev. Lett. {\bf 75}, (1995) 2847.

\bibitem{MezPar99}
  M. M\'ezard and G. Parisi,
  Phys. Rev. Lett. {\bf 82}, (1999) 747.

\bibitem{ColMezParVer99}
  B. Coluzzi, M. M\'ezard, G. Parisi and P. Verrocchio,
  J. Chem. Phys. {\bf 111}, (1999) 9039.

\bibitem{MezPar00}
  M. M\'ezard and G. Parisi,
  J. Phys.: Condens. Matter {\bf 12}, (2000) 6655.

\bibitem{ParZam05}
  G. Parisi F. Zamponi,
  J. Chem. Phys. {\bf 123}, (2005) 144501.

\bibitem{ParZam06}
  G. Parisi and F. Zamponi,
  J. Stat. Mech., (2006) P03017.

\bibitem{GroKraTarVio02}
  M. Grousson, V. Krakoviack, G. Tarjus and P. Viot,
  Phys. Rev. E {\bf 66}, (2002) 026126.

\bibitem{WesSchWol03}
  H. Westfahl Jr., J. Schmalian and P. Wolynes,
  Phys. Rev. B {\bf 68}, (2003) 134203.

\bibitem{MiyRei05}
  K. Miyazaki and D. Reichman,
  J. Phys. A.: Math. Gen. {\bf 38}, (2005) L343.

\bibitem{AndBirLef06}
  A. Andreanov, G. Biroli and A. Lefevre,
  J. Stat. Mech., (2006) P07008.

\bibitem{Gozzi83}
  E. Gozzi,
  Phys. Rev. D {\bf 28}, (1983) 1922.

\bibitem{Naketal83}
  Nakazato et al., Prog. Theo. Phys. {\bf 70}, 298 (1983).

\bibitem{BirCriprep}
  G. Biroli and A. Crisanti, in preparation.

\bibitem{risken}
  see for example
  H. Risken
  The Fokker-Planck Equation,
  (Springer-Verlag, Berlin 1989).

\bibitem{MarSigRos73}
  P.C. Martin, E. Siggia and H. Rose,
  Phys. Rev. A {\bf 8}, (1973) 423.

\bibitem{DeDom75}
  C. De Dominicis,
  Nuovo Cimento Lett. {\bf 12}, (1975) 567.

\bibitem{DeDom76}
  C. De Dominicis,
  J. Phys. (Paris) Colloq. {\bf 37}, (1976) C1.

\bibitem{Janssen76}
  H.K. Janssen,
  Z. f\"ur Phys. B {\bf 23}, (1976) 377.

\bibitem{BauJanWag76}
  R. Bausch, H.K. Janssen and H. Wagner,
  Z. f\"ur Phys. B {\bf 24}, (1976) 113.

\bibitem{DeDomPel78}
  C. De Dominicis and L. Peliti,
  Phys. Rev. B {\bf 18}, (1978) 353.

\bibitem{zinn}
  see for example
  J. Zinn-Justin,
  Quantum Field Theory and Critical Phenomena,
  (Oxford, Claredon 1996).

\bibitem{gotze}
  see for example
  W. G\"otze in 
  Liquids, Freezing and Glass Transition,
  J.P. Hansen, D. Levesque and J. Zinn-Justin (eds),
  (North-Holland, Amsterdam 1991), pag. 287.

\bibitem{lebellac}
  see for example,
  M. Le Bellac,
  Quantum and statistical field theory,
  (Oxford University Press, Oxford 1991)

\bibitem{DeDom63}
  C. De Dominicis,
  J. Math. Phys. {\bf 4}, (1963) 255.

\bibitem{CorJacTom74}
  J.M. Cornwall, R. Jackiw and E. Tomboulis,
  Phys. Rev. {\bf 10}, (1974) 2428.

\bibitem{Haymaker91}
  R.W. Haymaker,
  Riv. Nuovo Cimento {\bf 14}, (1991) 1.

\bibitem{ma}
  see for example
  S.K. Ma,
  Modern Theory of Critical Phenomena,
  Frontiers in physics 46,
  (Addison-Wesley, 1982).




\end{thebibliography}
\end{document}